\documentclass[prd,twocolumn,showpacs,nofootinbib,superscriptaddress]{revtex4}

\usepackage[usenames]{color}
\usepackage{amsfonts}
\usepackage{amsmath}
\usepackage{amssymb}
\usepackage{bm}
\usepackage{dcolumn}
\usepackage{epsfig}
\usepackage{graphicx}
\usepackage{graphics}
\usepackage[latin1]{inputenc}
\usepackage{latexsym}
\usepackage{rotating}
\usepackage{hyperref}
\usepackage{multirow}

\newcommand{\<}{\begin{equation}}
\newcommand{\?}{\end{equation}}

\newcommand{\cC}{\mathcal{C}}

\newcommand{\cG}{\mathcal{G}}

\newcommand{\cI}{\mathcal{I}}
\newcommand{\cL}{\mathcal{L}}
\newcommand{\cP}{\mathcal{P}}
\newcommand{\cQ}{\mathcal{Q}}
\newcommand{\cS}{\mathcal{S}}
\newcommand{\cU}{\mathcal{U}}

\newcommand{\GR}{\mathrm{GR}}
\newcommand{\Newt}{N} 

\newcommand{\eff}{\mathrm{eff}}

\newcommand{\Msolar}{\,M_{\odot}}

\begin{document}

\title{Maximum elastic deformations of relativistic stars}

\author{Nathan~K.~Johnson-McDaniel}

\affiliation{Institute for Gravitation and the Cosmos,
  Center for Particle and Gravitational Astrophysics,
  Department of Physics, The Pennsylvania State University, University
  Park, PA 16802, USA}
 \affiliation{Theoretisch-Physikalisches Institut,
  Friedrich-Schiller-Universit{\"a}t,
  Max-Wien-Platz 1,
  07743 Jena, Germany}

\author{Benjamin~J.~Owen}

\affiliation{Institute for Gravitation and the Cosmos,
  Center for Particle and Gravitational Astrophysics,
  Department of Physics, The Pennsylvania State University, University
  Park, PA 16802, USA}

\date{\today}

\begin{abstract}

We present a method for calculating the maximum elastic quadrupolar deformations
of relativistic stars, generalizing the previous Newtonian, Cowling
approximation integral given by [G.\ Ushomirsky \emph{et al.}, Mon.\ Not.\ R.\
Astron.\ Soc.\ {\bf{319}}, 902 (2000)]. (We also present a method for
Newtonian gravity with no Cowling approximation.) We apply these methods
to the $m=2$ quadrupoles most relevant for gravitational radiation in three cases:
crustal deformations, deformations of
crystalline cores of hadron--quark hybrid stars, and deformations of entirely
crystalline color superconducting quark stars.
In all cases, we find suppressions of the quadrupole due to relativity
compared to the Newtonian Cowling approximation, particularly for compact
stars.
For the crust these suppressions are up to a factor of $\sim 6$, for hybrid stars
they are up to $\sim 4$, and for solid quark stars they are at most $\sim 2$, with
slight enhancements instead for low mass stars.
We also explore ranges of masses and equations of state more than in previous
work, and find that for some parameters the maximum quadrupoles can still be very large.
Even with the relativistic suppressions, we find that $1.4\Msolar$ stars can
sustain crustal quadrupoles of $\text{a few}\times10^{39}\,\text{g cm}^2$ for
the SLy equation of state, or close to $10^{40}\,\text{g cm}^2$ for equations of
state that produce less compact stars. Solid quark stars of $1.4\Msolar$
can sustain quadrupoles of around $10^{44}\,\text{g cm}^2$.
Hybrid stars typically do not have solid cores at $1.4\Msolar$,
but the most massive ones ($\sim 2\Msolar$) can sustain quadrupoles of
$\text{a few}\times10^{41}\,\text{g cm}^2$ for typical microphysical parameters
and $\text{a few}\times10^{42}\,\text{g cm}^2$ for extreme ones.
All of these quadrupoles assume a breaking strain of $10^{-1}$ and can be
divided by $10^{45}\,\text{g cm}^2$ to yield the fiducial ``ellipticities''
quoted elsewhere.
\end{abstract}


\maketitle

\section{Introduction}

Shortly after the discovery of pulsars and the realization that they are
rotating neutron stars, deformations of rotating neutron stars were proposed as
sources of continuous gravitational radiation~\cite{Shklovskii1969,
Ostriker1969, Ferrari1969, Melosh1969}; see~\cite{Press1972} for an early
review.
Searches for such radiation are an ongoing concern of the LIGO and Virgo
gravitational wave detectors~\cite{LIGO_psrs2010, LIGO_CasA, LIGO_Vela}; see~\cite{Owen2009, Pitkin, Astone} for recent
reviews.
It is thus of
great interest to know the maximum
quadrupolar deformation that a neutron star could sustain, in order to motivate
further searches and help interpret upper limits or detections.
In the case of elastic (as opposed to magnetic) deformations, the main factor
influencing the answer is whether the neutron star contains particles more
exotic than neutrons~\cite{OwenPRL, Owen2009}.
However, the structure of the star also plays an important role.

While there are
relativistic calculations of the quadrupole deformations due to magnetic
fields (e.g.,~\cite{IS,CFG,YKS, FR,CR}), all the computations involving elastic
deformations have used Newtonian gravity. Moreover, all but two of these computations
have used the integral expression obtained in the Cowling approximation (i.e.,
neglecting the self-gravity of the perturbation) by
Ushomirsky, Cutler, and
Bildsten (UCB)~\cite{UCB}; see \cite{OwenPRL, Lin, KS, Horowitz}.
Haskell, Jones, and Andersson (HJA)~\cite{HJA} dropped the Cowling approximation using
a somewhat different formalism than UCB's; there is a further application of
their results in~\cite{Haskelletal}.

We improve these treatments by generalizing the UCB integral to relativistic
gravity with no Cowling approximation. We also provide a
similar generalization for the Newtonian no-Cowling case, as a warm-up. In
addition to providing a simpler formalism for performing computations than the
more general Newtonian gravity treatment in HJA, the integrals we obtain allow us to verify that a
maximal uniform
strain continues to yield the maximum quadrupole deformation in the Newtonian
and relativistic no-Cowling cases. (UCB showed this to be true for an arbitrary equation of state
in the Newtonian Cowling approximation case; we are able to verify that it is
true in the more general cases for each background stellar model we consider.) 

We then apply our calculation to the standard case of quadrupoles supported by
shearing the lattice of nuclei in the crust, as well as the cases where the
quadrupole is supported by the hadron--quark mixed phase lattice in the core, or a crystalline color superconducting phase throughout a solid strange quark star. For the crustal
quadrupoles, we calculate the shear modulus following HJA, using the equation
of state (EOS) and composition results of Douchin and Haensel~\cite{DH} and
the effective shear modulus calculated by Ogata and Ichimaru~\cite{OI}. (There are recent
improvements to the Ogata and Ichimaru result~\cite{HH, Baiko,BaikoCPP}, but these only reduce their
shear modulus by $<10\%$.) For the hadron--quark mixed phase, we use our recent calculations of the EOS and
shear modulus~\cite{J-MO1} for a variety of parameters. (We also consider the range of surface tensions for
which the mixed phase is favored.) For crystalline quark matter, we use the shear modulus calculated by Mannarelli, Rajagopal, and Sharma~\cite{MRS}, and the EOS given by
Kurkela, Romatschke, and Vuorinen~\cite{KRV}.

In all cases, we use a breaking strain of $0.1$, comparable to that
calculated by Horowitz and Kadau~\cite{HK} using molecular dynamics
simulations. (Hoffman and Heyl~\cite{HoHe} have recently obtained
very similar values over more of parameter space.) This result is directly applicable to the crustal lattice, at least for the outer crust, above neutron drip (though see Chugunov and
Horowitz~\cite{ChHo} for caveats). We also
feel justified in applying it to the inner crust, as well as to the mixed phase and crystalline quark matter, since the primary source of
the high breaking strain appears to be the system's large pressure.
But one can apply our results to any breaking strain using the linear scaling of
the maximum quadrupole with breaking strain.

In our general relativistic calculation, we use the relativistic theory of elasticity given by Carter and
Quintana~\cite{CQ} and placed in a more modern guise by
Karlovini and Samuelsson~\cite{KaSaI}.
However, all we need from it is the relativistic form of the elastic stress-energy tensor, which can be obtained by simple
covariance arguments, as noted by Schumaker and Thorne~\cite{ST}. We also use the standard
\citet{TC} Regge-Wheeler gauge~\cite{RW} formalism for perturbations
of static relativistic stars, following
Hinderer's recent calculation~\cite{Hinderer} of the quadrupole moment of a
tidally deformed relativistic star (first discussed in~\citet{FH}), and the classic calculation by \citet{Ipser}.

Even though we are interested in the gravitational radiation emitted by
rotating stars, it is sufficient for us to calculate the static quadrupole deformation.
As discussed by
Ipser~\cite{Ipser}, and then proved for more general situations by
Thorne~\cite{Thorne}, this static quadrupole (obtained
from the asymptotic
form of the metric) can be inserted into the quadrupole formula to obtain the emitted
gravitational radiation in the fully relativistic,
slow-motion limit. [This
approximation has uncontrolled remainders of order $(\omega/\omega_K)^2$, where
$\omega$ and $\omega_K$ are the star's angular velocity and its maximum---i.e.,
Kepler---angular velocity, respectively.
This ratio is $\lesssim 10^{-2}$ for the pulsars for which LIGO has been able
to beat the spin-down limit~\cite{LIGO_psrs2010}.]

We shall generally show the gravitational constant $G$ and speed of light $c$
explicitly, though we shall take $G = c =1$ in most of Sec.~\ref{GR}, only restoring them in our
final expressions.
The relativistic calculation was aided by use of the computer algebra
system {\sc{Maple}}
and the associated tensor manipulation package {\sc{GRTensorII}}~\cite{GRTensor}.
We used {\sc{Mathematica}}~7 to perform numerical computations.

The paper is structured as follows: In Sec.~\ref{Newt}, we review UCB's formalism
and extend it by introducing a Green function to compute the maximum Newtonian quadrupole deformation without
making the Cowling approximation. In Sec.~\ref{GR}, we further generalize to
the fully relativistic case, and compare the various approximations for the maximum quadrupole.
In Sec.~\ref{results2}, we show the maximum quadrupoles for three different cases: first crustal quadrupoles,
then hadron--quark hybrid quadrupoles, and finally solid strange quark star quadrupoles. We also describe
the modifications to our formalism needed to treat solid strange quark stars. We discuss all these
results in Sec.~\ref{discussion}, and summarize and conclude in
Sec.~\ref{concl2}.
In the Appendix, we show that the mixed phase is favored by global energy arguments even for surface tensions large enough that it is disfavored by local energy arguments.

\section{Newtonian calculation of the maximum quadrupole}
\label{Newt}

We first demonstrate how to compute the maximum Newtonian quadrupole without making the Cowling approximation. This provides
a warm-up before we tackle the full relativistic case, and also allows us to verify some of the
statements made by UCB and HJA. We use
the basic formalism of UCB, modeling the star as nonrotating, with the
stress-energy tensor of a perfect fluid plus
shear terms, and treating the shear contributions as a first-order
perturbation of hydrostatic equilibrium.
This perturbative treatment should be quite a good approximation: The maximum shear stress to energy density ratio we consider in the crustal and hybrid star cases is $\lesssim 0.05\%$ (and the maximum shear stress to pressure ratio is
$\lesssim 0.3\%$). (Here we have taken the shear stress to
be $\mu\bar{\sigma}_\mathrm{max}$, which is good up to factors of order unity.) And even in the case of solid strange quark stars, the
maximum shear stress to energy density ratio is still only at most $\sim 0.2\%$.
[We have already discussed the effects of rotation in the relativistic case, above; UCB note at
the beginning of their Sec.~4 that rotation also only modifies the perturbative Newtonian results for the static deformations we and they consider at the $O([\omega/\omega_K]^2)$ level.] 

It is convenient to start by writing the quadrupole moment
in terms of the surface value of the perturbation to the star's
Newtonian potential. We start from UCB's definition of
\<
\label{Q22}
Q_{22} := \int_0^\infty\delta\rho(r)r^4dr
\?
[where the (Eulerian) density perturbation $\delta\rho$ and all similar
perturbed quantities have only an $l = m = 2$ spherical harmonic component].
[Note that this
quadrupole moment differs by an overall constant from the one defined by
Thorne~\cite{Thorne}---e.g., his Eq.~(5.27a).] We then recall that
the perturbed Poisson equation for the $l = 2$ part of the perturbed
gravitational potential is
\<\label{Poisson}
(\triangle_2\delta\Phi)(r) := \frac{1}{r^2}[r^2\delta\Phi'(r)]' - \frac{6}{r^2}\delta\Phi(r) = 4\pi G\delta\rho
\?
($\triangle_2$ is the $l = 2$ radial part of the Laplacian), with boundary conditions of
\<
\label{PoissonBCs}
\delta\Phi(0) = 0, \qquad R\delta\Phi'(R) = -3\delta\Phi(R),
\?
where $R$ is the radial coordinate of the star's surface.
[See, e.g., Eqs.~(2.15) and (2.16) in \cite{LMO}---their $\Phi_{22}$ is our
$\delta\Phi$. Note also that the primes denote derivatives with respect to $r$. Additionally, we shall
continue to be inconsistent with our inclusion of the functional dependence of quantities---e.g., $\delta\rho$
depends upon $r$, even though we do not always indicate this explicitly. We will eventually stop displaying  $\delta\Phi$'s 
explicit functional dependence on $r$, for instance.]
If we now substitute Eq.~\eqref{Poisson} into Eq.~\eqref{Q22} and integrate by
parts using the boundary conditions~\eqref{PoissonBCs}, we obtain
\<\label{Q22N}
Q_{22} = -\frac{5R^3}{4\pi G}\delta\Phi(R).
\?
This
sort of expression is more commonly seen in the relativistic case, where it is
necessary to obtain the quadrupole in this manner by looking at the
perturbation's asymptotic behavior---see the discussion in Sec.~\ref{GR}.

We now wish to obtain an equation for $\delta\Phi$ in terms of the shear
stresses. We follow UCB in decomposing the perturbed stress tensor as
[see their Eqs.~(59) and (61)]
\<\label{deltatau}
\begin{split}
\delta\tau_{ab} &= -\delta pY_{lm}g_{ab} + t_{rr}Y_{lm}(\hat{r}_a\hat{r}_b - e_{ab}/2) + 
t_{r\perp}f_{ab}\\
&\quad + t_\Lambda(\Lambda_{ab} + Y_{lm}e_{ab}/2).
\end{split}
\? 
Here $\delta p$ is the (Eulerian) pressure perturbation; $Y_{lm}$ is a spherical
harmonic; $\hat{r}_a$ is the radial unit vector; $t_{rr}$, $t_{r\perp}$, and
$t_\Lambda$ are the components of the shear stresses; and
$g_{ab}$ denotes the metric of flat, $3$-dimensional Euclidean space.
(Following UCB, we will generally write out $l$ and $m$ explicitly, even
though we only consider $l = m = 2$ here.)
Also [Eqs.~(40) in UCB],
\begin{subequations}
\begin{align}
e_{ab} &:= g_{ab} - \hat{r}_a\hat{r}_b,
\\
f_{ab} &:= 2r\hat{r}_{(a}\nabla_{b)}Y_{lm}/\beta,
\\
\beta &:= \sqrt{l(l+1)} = \sqrt{6},
\\
\label{Lambda}
\Lambda_{ab} &:= r^2\nabla_a\nabla_bY_{lm}/\beta^2 + f_{ab}/\beta.
\end{align}
\end{subequations}
(We have corrected the dropped factor of $\beta^{-1}$ multiplying $f_{ab}$ in
UCB's definition of $\Lambda_{ab}$---this was also noticed by HJA.)
We also have
\begin{equation}
\label{ts}
t_{ab} = 2\mu\sigma_{ab},
\end{equation}
where $\mu$ is the shear modulus and $\sigma_{ab}$ is the strain tensor.
(This is a factor-of-$2$ correction to the expression in UCB, as noted in
\cite{OwenPRL}.)
Now, a convenient expression can be obtained from the perturbed
equation of hydrostatic equilibrium
\<\label{hydro}
\nabla^a\delta\tau_{ab} = \delta\rho g(r)\hat{r}_b + \rho\nabla_b\delta\Phi
\?
($\nabla_a$ denotes the flat-space covariant derivative), by
substituting for $\delta\rho$ using the Poisson equation~\eqref{Poisson}
and projecting along $\hat{r}^b$, yielding
\<\label{deltaPhieq1}
\begin{split}
\frac{\triangle_2\delta\Phi}{4\pi G} + \frac{\rho}{g(r)}\delta\Phi' &= \frac{\hat{r}^b\nabla^a\delta\tau_{ab}}{g(r)}\\
&= \frac{1}{g(r)}\left[-\delta p' + t_{rr}' + \frac{3}{r}t_{rr} - \frac{\beta}{r}t_{r\perp}\right].
\end{split}
\?
We then project Eq.~\eqref{hydro} along $\nabla^bY_{lm}$ to express $\delta p$ in terms of the shear stresses $t_{rr}$, $t_{r\perp}$, and $t_\Lambda$,
along with $\rho$ and $\delta\Phi$, giving
\<
\delta p = -\rho\delta\Phi - \frac{t_{rr}}{2}  + \frac{r}{\beta}t_{r\perp}' + \frac{3}{\beta}t_{r\perp} + \left(\frac{1}{\beta^2} - \frac{1}{2}\right)t_\Lambda.
\?
Substituting this into Eq.~\eqref{deltaPhieq1}, we thus obtain
\<\label{deltaPhieq2}
\begin{split}
\triangle_2\delta\Phi - \frac{4\pi G}{g(r)}\rho'\delta\Phi &=
\frac{4\pi G}{g(r)}\biggl[\frac{3}{2}t_{rr}' - \frac{4}{\beta}t_{r\perp}' - \frac{r}{\beta}t_{r\perp}''\\
&\quad - \left(\frac{1}{\beta^2} - \frac{1}{2}\right)t_\Lambda' + \frac{3}{r}t_{rr} - \frac{\beta}{r}t_{r\perp}\biggr].
\end{split}
\?

We now wish to obtain an integral expression for $Q_{22}$ that generalizes
UCB's Eq.~(64) to the case where we do not make the Cowling approximation.
We shall do this by obtaining the Green function for the left-hand side of
Eq.~\eqref{deltaPhieq2} and then integrating by parts.
We will be able to
discard all of the boundary terms, since the
stresses vanish at the star's surface (we assume that the shear modulus vanishes there) and the integrand vanishes at the star's center. We can obtain the
Green function using the standard Sturm-Liouville expression in terms of the solutions of the
homogeneous equation [e.g., Eq.~(10.103) in Arfken and Weber~\cite{AW}~].
We obtain the appropriate solution to the homogeneous equation numerically
for a given background stellar model (EOS and mass). The equation for the Green function is
[multiplying the left-hand side of Eq.~\eqref{deltaPhieq2} by $r^2$ to improve its regularity]
\<\label{Leq}
\begin{split}
(\cL_N\cG)(r,\bar{r}) &:= \frac{\partial}{\partial r}\left[r^2\frac{\partial}{\partial r}\cG(r,\bar{r})\right] - \left[6 + \frac{4\pi G r^2}{g(r)}\rho'\right]\cG(r,\bar{r})\\
&\,= \delta(r - \bar{r})
\end{split}
\?
[$\delta(r - \bar{r})$ is the Dirac delta function], with boundary conditions (at the star's center and surface) of
\<\label{Newt_BC}
\cG(0,\bar{r})=0, \qquad R\partial_1\cG(R,\bar{r}) = -3\cG(R,\bar{r}),
\?
where $\partial_1$ denotes a partial derivative taken with respect to the
first ``slot'' of the function.

If we then write [using Eq.~\eqref{Q22N}, the factor of $r^2$ from the Green function equation~\eqref{Leq}, and the
prefactor on the right-hand side of Eq.~\eqref{deltaPhieq2}]
\begin{equation}\label{GN}
G_\Newt(r) := -5R^3r^2\cG(R,r)/g(r),
\end{equation}
we have
\begin{widetext}
\<\label{Q22N1}
\begin{split}
Q^\Newt_{22} &= \int_0^RG_\Newt(r)\left[\frac{3}{2}t_{rr}' - \frac{4}{\beta}t_{r\perp}' - \frac{r}{\beta}t_{r\perp}'' -
\left(\frac{1}{\beta^2} - \frac{1}{2}\right)t_\Lambda' + \frac{3}{r}t_{rr} - \frac{\beta}{r}t_{r\perp}\right]dr\\
&= -\int_0^R\biggl\{\left[\frac{3}{2}G_\Newt'(r) - \frac{3}{r}G_\Newt(r)\right]t_{rr} + \left[\frac{r}{\beta}G_\Newt''(r) - \frac{2}{\beta}G_\Newt'(r) + \frac{\beta}{r}G_\Newt(r)\right]t_{r\perp}
+ \left(\frac{1}{2} - \frac{1}{\beta^2}\right)G_\Newt'(r)t_\Lambda\biggr\}dr.
\end{split}
\?
\end{widetext}
We have freely integrated by parts in obtaining the second expression, noting that the boundary terms are zero since $G_\Newt(r)$ vanishes sufficiently rapidly as $r \to 0$ and
the stresses are zero at the surface of the star (since we assume that the shear modulus vanishes at the star's surface).\footnote{We shall treat the case where the stresses do \emph{not} vanish at the surface of the star when we consider solid strange quark stars in Sec.~\ref{SQM_computation}. Also, note that HJA claim that UCB's
expression does not include distributional contributions due to sudden changes
in the shear modulus. This is not the case---these are included due to UCB's
integration by parts (cf.\ the definition of the distributional derivative). All that the UCB derivation requires is,
e.g.,  that the shear modulus vanish outside of the crust, not that
it do so continuously.}
This reduces to UCB's Eq.~(64) if we take the Cowling approximation
\begin{equation}
\label{CowlingGN}
G_\Newt(r) \to r^4/g(r),
\end{equation}
corresponding to dropping the second term on the left-hand side of
Eq.~\eqref{deltaPhieq2}.

To obtain an analogue of the expression for the maximum quadrupole
given in Eq.~(5) of Owen~\cite{OwenPRL}, we note that
UCB's argument about maximum uniform strain leading to the maximum quadrupole
still holds here for the stars we consider, since the coefficients of the stress
components in the integrand are all uniformly positive. (We have checked this numerically for each background
stellar model we consider.)
The strain tensor components are
\begin{subequations}
\label{sNewt}
\begin{align}
\label{srr}
\sigma_{rr} &= (32\pi/15)^{1/2}\bar{\sigma}_\mathrm{max},
\\
\label{srp}
\sigma_{r\perp} &= (3/2)^{1/2}\sigma_{rr},
\\
\label{sL}
\sigma_\Lambda &= 3\sigma_{rr}
\end{align}
\end{subequations}
in the case where the
star is maximally (and uniformly) strained---see Eqs.~(67) in UCB. The
breaking strain $\bar{\sigma}_\mathrm{max}$ is given by the von Mises
expression,
\begin{equation}
\label{vonMises}
\sigma_{ab}\sigma^{ab} = 2\bar{\sigma}_\mathrm{max}^2.
\end{equation}
It thus
corresponds to assuming that the lattice yields when it has stored a certain maximum
energy density. We then have
\<\label{Q22N2}
\frac{|Q_{22}^{\mathrm{max}, N}|}{\bar{\sigma}_\mathrm{max}} = \sqrt{\frac{32\pi}{15}}\int_0^R\mu(r)\left[rG_\Newt''(r) + 3G_\Newt'(r)\right]dr.
\?
This reduces to Eq.~(5) in Owen~\cite{OwenPRL} if we use the Cowling
approximation~\eqref{CowlingGN}.

Note that there is no direct contribution from $\rho'$ to $G_\Newt''$ in the
no-Cowling case, despite what one might expect from Eq.~\eqref{Leq}:  Writing
$\bar{\cG}(r) := \cG(R,r)$ for notational simplicity, the $\rho'$ contribution from
\<
\bar{\cG}''(r) = (2/r)\bar{\cG}'(r) + 
[6/r^2 + 4\pi G\rho'(r)/g(r)]\bar{\cG}(r)
\?
is exactly canceled by one from 
\<
g''(r) = 6Gm(r)/r^4 - 8\pi G\rho(r)/r + 4\pi G\rho'(r)
\?
in
\<
\begin{split}
G''_\Newt(r) &= -5R^3r^2[\bar{\cG}''(r)/g(r) - \bar{\cG}(r)g''(r)/\{g(r)\}^2\\
&\quad + \{\text{terms with no }\rho'\}].
\end{split}
\?
However, there is a direct contribution from
$\rho'$ to $G_\Newt''$ (via $g''$) if we make the Cowling
approximation [Eq.~\eqref{CowlingGN}].
We shall see that this leads to a
significant
difference in the resulting contributions to the quadrupole moment from regions
of the star surrounding a sudden change in density (e.g., near the crust-core
interface, which will be relevant for the quadrupoles supported by crustal
elasticity considered by UCB and others).

Numerically, we compute $G_\Newt$ using the standard expression for the Green function in terms of the
two independent solutions to the homogeneous equation [see, e.g., Eq.~(10.103) in Arfken and Weber~\cite{AW}~]. Since we are solely interested in the Green function evaluated at the star's surface, we can
eliminate one of the homogeneous solutions using the boundary conditions there, and only consider the
homogeneous solution that is regular at the origin, which we call $F$.
In terms of $F$, the Green function is given by
\<\label{cG_F}
\cG(R,r) = -\frac{F(r)}{3RF(R) + R^2F'(R)}.
\?
We thus solve $\cL_N F = 0$ [with the operator $\cL_N$ given by Eq.~\eqref{Leq}] with the boundary conditions
$F(r_0) = 1$ and $F'(r_0) = 2/r_0$, where $r_0$ is the small inner radius used in the solution of the
OV equations, as discussed at the end of Sec.~\ref{GR}. [These boundary conditions come from
regularity at the origin, which implies that $F(r) = O(r^2)$ there.]

Our Green function method for obtaining the maximum quadrupole numerically may
seem more complicated than existing methods because it introduces extra steps.
However this method is ideal for showing that maximum stress gives the maximum
quadrupole and for seeing how much stresses at different radii contribute to
the total quadrupole.
It also appears to be the simplest way of dealing with any potential
distributional contributions from the derivatives of the shear modulus, since
they are automatically taken care of by the integration by parts.

\section{General relativistic calculation of the maximum quadrupole}
\label{GR}

Here we compute the maximum quadrupole moment in general relativity, using the Regge-Wheeler
gauge~\cite{RW} relativistic stellar perturbation theory developed by \citet{TC}, as in
the similar calculation of the tidal Love number of a relativistic star by
\citet{Hinderer}.
We
start by writing down the line element corresponding to a static, even-parity,
$l = 2$ first-order perturbation of a static,
spherical, relativistic star in the Regge-Wheeler gauge [cf.\ Eq.~(14) in Hinderer~\cite{Hinderer}~]:
\<\label{ds2}
\begin{split}
ds^2 &= -[1 + H_0(r)Y_{lm}]f(r)dt^2 + [1 + H_2(r)Y_{lm}]h(r)dr^2\\
&\quad + [1 + K(r)Y_{lm}]r^2(d\theta^2 + \sin^2\theta d\phi^2).
\end{split}
\?
Here we have used the notation of Wald~\cite{Wald} for the background, so that $f$ and $h$ are the standard
Schwarzschild functions for the unperturbed star, with $f = e^{2\phi}$, where
\<
\phi'(r) = \frac{m(r) + 4\pi r^3 p}{r[r - 2m(r)]},
\?
with $\phi(R) = \log(1 - 2M/R)/2$, and
\<
h(r) = \left[1 - \frac{2m(r)}{r}\right]^{-1}.
\?
In these expressions,
\begin{equation}
m(r) := 4\pi \int_0^r \rho(\bar{r})\bar{r}^2d\bar{r}.
\end{equation}
Also, recall that we write our spherical harmonics in terms of $l$ and $m$, following
UCB, even though we specialized to $l = m = 2$, and that we are now taking $G = c = 1$.

The metric perturbation is determined by $H_0$, $H_2$, and $K$, which here are sourced by the
perturbation to the star's stress-energy tensor. The appropriate stress-energy
tensor can be obtained directly
from the standard Newtonian expression~\eqref{deltatau} by simple covariance arguments, as in
Schumaker and Thorne~\cite{ST}, or from the detailed relativistic elasticity
theory of Carter and Quintana~\cite{CQ} [see their Eq.~(6.19); this is also given in Eq.~(128) of
Karlovini and Samuelsson~\cite{KaSaI}~]. All
we really need for our purposes is to note that the shear contribution is
tracefree with respect to the background metric, so that we can use the obvious
covariant generalization of the decomposition given by UCB,\footnote{Of course, this assumes that it is
possible to obtain \emph{any} symmetric tracefree tensor from the detailed relativistic expression, but---as would be expected (and can easily be seen from the expressions)---this is indeed the case, at least if one only works to first order in the perturbation, as we
do here.
Also, it is instructive to note that we do not need to know the specifics of the matter displacements that generate
the quadrupoles we consider, only that there is a tracefree contribution to the star's
stress-energy tensor whose maximum value is given by the material's shear
modulus and von Mises breaking strain.} yielding 
\<\label{deltaT}
\begin{split}
\delta T_{ab} &= [\delta\rho\hat{t}_a\hat{t}_b +  \delta p(g_{ab} + \hat{t}_a\hat{t}_b) - 
t_{rr}(\hat{r}_a\hat{r}_b - q_{ab}/2)]Y_{lm}\\
&\quad - 
t_{r\perp}f_{ab} - t_\Lambda(\tilde{\Lambda}_{ab} + h^{1/2}Y_{lm}q_{ab}/2),
\end{split}
\?
with the full stress-energy tensor given by
\<
T_a{}^b = \rho\hat{t}_a\hat{t}^b + p(\delta_a{}^b + \hat{t}_a\hat{t}^b) + \delta T_a{}^b.
\?
Here, indices now run over all four spacetime dimensions and $g_{ab}$ denotes
the background (spacetime) metric (which we use to raise and lower indices).
Additionally, we have introduced the background temporal and radial unit vectors $\hat{t}_a$ and $\hat{r}_a$;
$q_{ab}$ is the induced metric on the unit $2$-sphere;
$f_{ab} := 2r\hat{r}_{(a}\nabla_{b)}Y_{lm}/\beta$;
and
$\tilde{\Lambda}_{ab} :=  r^2h^{1/2}\nabla_a\nabla_bY_{lm}/\beta^2 + f_{ab}/\beta$.
Here $\hat{r}_a$ and $\nabla_a$ now have their curved-space meanings.

Our $\tilde{\Lambda}_{ab}$
differs from the Newtonian $\Lambda_{ab}$ [from UCB, given in our Eq.~\eqref{Lambda}] due to the insertion of $h^{1/2}$. This insertion is necessary
for $\tilde{\Lambda}_{ab}$ to be transverse and orthogonal to $f_{ab}$ (with respect to the background spacetime metric).
The same logic leads to the introduction of the factor of $h^{1/2}$ multiplying $q_{ab}$ in the $t_\Lambda$ term in
Eq.~\eqref{deltaT}; it is there so that the $t_\Lambda$ term is orthogonal to the $t_{rr}$ term. We
have used UCB's convention
for the relative sign between the perfect fluid and shear portions of the stress-energy tensor, though we
have reversed the overall sign. (However, we used the UCB convention proper
in Sec.~\ref{Newt}.)
The factor of $h^{1/2}$ in the coefficient of $t_\Lambda$ leads to a factor
of $h^{-1}$ in the
strain $\sigma_\Lambda$ that corresponds to the von Mises
breaking strain~\eqref{vonMises}.
We thus replace the Newtonian Eq.~\eqref{sL} with
\begin{equation}
\label{sLGR}
\sigma_\Lambda = 3\sigma_{rr}/h,
\end{equation}
leaving Eqs.~\eqref{srr} and~\eqref{srp} unchanged.

One can now obtain an equation for $H_0$ from the perturbed Einstein equations,
as in Ipser~\cite{Ipser}. (The other two metric functions, $H_2$ and $K$, can be expressed in terms of $H_0$;
these expressions are given by Ipser.) The concordance for notation
is $\nu = 2\phi$, $e^\nu = f$, $\lambda = 2\psi$, $e^\lambda = h$, $\rho_1 = -\delta\rho$, $p_1 = -\delta p$,
$\mathfrak{P}_2 = t_{rr}$, $\mathfrak{Q}_1 = h^{1/2}t_{r\perp}/\beta$, and $\mathfrak{S} = h^{1/2}t_\Lambda/\beta^2$. Additionally, Ipser's $H_0$ is the
negative of ours. The relevant result is given in Ipser's Eqs.~(27)--(28), and
is (in our notation)
\<
H_0'' + \left(\frac{2}{r} + \phi' -\psi'\right)H_0' + \cP(r)H_0= 8\pi h^{1/2}\cS(r),
\?
where
\<\label{cP}
\cP(r) := 2\phi'' + 2\phi'\left(\frac{3}{r} - \phi' -\psi'\right) + \frac{2\psi'}{r} - \frac{\beta^2}{r^2}h
\?
and
\<\label{cS}
\begin{split}
\cS(r) &:=
h^{1/2}(\delta\rho + \delta p - t_{rr}) +
2\biggl\{(3 - r\phi')\frac{t_{r\perp}}{\beta} + r\frac{t_{r\perp}'}{\beta}\\
&\,\quad + [r^2\phi'' + r\phi'(5 - r\phi') + r\psi' -\beta^2h/2 + 1]\frac{t_\Lambda}{\beta^2}\\
&\,\quad + r^2\phi'\frac{t_\Lambda'}{\beta^2}\biggr\}\\
&=: h^{1/2}(\delta\rho + \delta p) + \cS_{[t]}(r).
\end{split}
\?
Here we have defined $\psi := (1/2)\log h$ and written $\cS_{[t]}$
for the contributions from shear stresses. (The ``$=:$'' notation implies that the quantity
being defined is on the right-hand side of the equality.)

We now wish to eliminate $\delta\rho$ and $\delta p$ in favor of the shear stresses, as in the Newtonian calculation. We use the same projections of stress-energy conservation as in the Newtonian case (projecting onto the quantities defined by
the background spacetime, for simplicity) along with the Oppenheimer-Volkov
(OV) equations, giving
\<
\begin{split}
\delta\rho + \delta p &= \frac{1}{\phi'}\biggl[-\frac{H_0'}{2}(\rho + p)
-\delta p' + t_{rr}' + \left(\frac{3}{r} + \phi'\right)t_{rr}\\
&\quad - \frac{\beta}{r}h^{1/2}t_{r\perp}\biggr]
\end{split}
\?
and
\<
\begin{split}
\delta p &= -\frac{H_0}{2}(\rho + p) - \frac{t_{rr}}{2} + \frac{1}{\beta h^{1/2}}\left[(3 + r\phi')t_{r\perp} + rt_{r\perp}'\right]\\
&\quad + h^{1/2}\left(\frac{1}{\beta^2} - \frac{1}{2}\right)t_\Lambda.
\end{split}
\?
Using the second expression to substitute for $\delta p'$ in the first, we have
\<\label{drplusdp}
\begin{split}
\delta\rho + \delta p &= \frac{1}{\phi'}\biggl\{\frac{H_0}{2}(\rho' + p') + \left[\frac{3}{r} + \phi'\right]t_{rr} + \frac{3}{2}t_{rr}'\\
&\quad - \frac{1}{\beta h^{1/2}}\biggl[\left(\frac{\beta^2 h}{r} + \phi' + r\phi'' - \psi'[3 + r\phi'] \right)t_{r\perp}\\
&\quad + (4 + r[\phi' - \psi'])t_{r\perp}' + rt_{r\perp}''\biggr] + 
\left(\frac{1}{2} - \frac{1}{\beta^2}\right)\\
&\quad\times h^{1/2}(\psi't_\Lambda + t_\Lambda')\biggr\}\\
&=: \frac{H_0}{2\phi'}(\rho' + p') + \frac{\cS_{[\delta\rho,\delta p]}(r)}{\phi'}.
\end{split}
\?
The equation for $H_0$ thus becomes
\<\label{cLGR}
\begin{split}
(\cL_\GR H_0)(r) &:= H_0'' + \left(\frac{2}{r} + \phi' -\psi'\right)H_0'\\
&\,\quad + \left[\cP(r) - 4\pi h\frac{\rho' + p'}{\phi'}\right]H_0\\
&\,= 8\pi h^{1/2}[h^{1/2}\cS_{[\delta\rho,\delta p]}(r)/\phi' + \cS_{[t]}(r)].
\end{split}
\?
[$\cP(r)$ and $\cS_{[t]}(r)$ are given in Eqs.~\eqref{cP} and~\eqref{cS}, respectively.]
As expected, this
reduces to Eq.~\eqref{deltaPhieq2} in the Newtonian
limit [where we have $H_0 \to 2\delta\Phi$ and $\phi' \to g(r)$].

We now want to write the equation for $H_0$ in Sturm-Liouville form in order to obtain its Green
function easily. To do this, we note that the appropriate ``integrating factor'' (for the first
two terms) is $r^2(f/h)^{1/2}$, which gives
\begin{multline}
\label{H0_S-L}
[r^2(f/h)^{1/2}H_0']' + r^2(f/h)^{1/2}\left[\cP(r) - 4\pi h\frac{\rho' + p'}{\phi'}\right]H_0\\
= 8\pi r^2f^{1/2}[h^{1/2}\cS_{[\delta\rho,\delta p]}(r)/\phi' + \cS_{[t]}(r)].
\end{multline}
We also need the boundary conditions, which are given by matching $H_0$ onto a
vacuum solution at the surface of the star. The vacuum
solution that is regular at infinity is given by Eq.~(20) in Hinderer~\cite{Hinderer} with $c_2 = 0$,
viz.,
\<\label{H0_BC}
\begin{split}
H_0(R) &= c_1\biggl[\left(\frac{2}{\cC} - 1\right)\frac{\cC^2/2 + 3\cC - 3}{1 - \cC}\\
&\quad + \frac{6}{\cC}\left(1 - \frac{1}{\cC}\right)\log\left(1 - \cC\right)\biggr],
\end{split}
\?
where we have evaluated this at the star's surface ($r = R$) and
defined the star's compactness
\<
\label{compactness}
\cC := 2GM/Rc^2
\?
(now returning to showing factors of $G$ and $c$ explicitly).
We require that $H_0$
and $H_0'$ be continuous at the star's surface. The value of $c_1$ obtained from
this matching of the internal and external solutions gives us 
the quadrupole moment. If we use the quadrupole moment amplitude that reduces
to the UCB integral [given in our Eq.~\eqref{Q22}] in the Newtonian limit, we have
\<\label{Q22rel}
Q_{22} = \frac{G^2}{c^4}\frac{M^3c_1}{\pi}.
\?
[This expression comes from inserting a pure $l = m = 2$ density perturbation
into Eq.~(2) in Hinderer~\cite{Hinderer}, contracting the free indices with
unit position vectors, performing the angular integral, for
which the expressions in Thorne~\cite{Thorne} are useful,
and noting that the result is $(8\pi/15)Y_{22}$ times
our Eq.~\eqref{Q22}. The
given result then follows immediately from Hinderer's Eqs.~(7), (9), and (22);
we reverse the overall sign since we have reversed the UCB sign convention for the
stress-energy tensor.]

We then have a Green function for $Q_{22}$ of 
\<\label{cG_GR}
\begin{split}
\cG_\GR(R,r) &= \left(\frac{2GM}{c^2}\right)^3\left(1 - \frac{2GM}{Rc^2}\right)^{-1}\\
&\quad\times\frac{\cU(r)}{c^2R^2[\cU'(R)H_0(R) - \cU(R)H_0'(R)]}
\end{split}
\?
(including the overall factor of $8\pi G/c^4$ that
multiplies the source).
Here $\cU$ is given by $\cL_\GR\cU = 0$ [$\cL_\GR$ is given in
Eq.~\eqref{cLGR}], with boundary conditions $\cU(r_0) = 1$ and $\cU'(r_0) = 2/r_0$.
[Compare Eq.~(10.103) in Arfken and Weber~\cite{AW}, as well as our Newtonian
version above.]
Additionally, $H_0(R)$ and $H_0'(R)$ are given by the boundary conditions~\eqref{H0_BC}
with $c_1 \to 1$. [One obtains this expression by first computing the Green function
for $H_0(R)$ following Arfken and Weber, then dividing through by the quantity in brackets in Eq.~\eqref{H0_BC}
to obtain $c_1$, and finally using Eq.~\eqref{Q22rel} to obtain $Q_{22}$. We have also noted that
$1/f \to h \to 1/(1 - 2GM/Rc^2)$ at the star's surface.]
We thus define, for notational simplicity, two relativistic generalizations
of $G_\Newt(r)$: One,
\<\label{GGR}
G_\GR(r) := \frac{r^2(fh)^{1/2}\cG_\GR(R,r)}{\phi'},
\?
for
the contributions from $\cS_{[\delta\rho,\delta p]}$, and one,
\<\label{GbGR}
\bar{G}_\GR(r) := r^2f^{1/2} \cG_\GR(R,r),
\?
for the contributions
from $\cS_{[t]}$. 

With these definitions, the integral expression for the quadrupole in terms
of the stresses and the
structure of the background star is
\<\label{Q22GR}
\begin{split}
Q_{22} &= \int_0^R\left[G_\GR(r)\cS_{[\delta\rho,\delta p]}(r)+\bar{G}_\GR(r)\cS_{[t]}(r)\right]dr\\
&= \int_0^R(\cC_{rr}t_{rr} + \cC_{t\perp}t_{r\perp} + \cC_\Lambda t_\Lambda)dr,
\end{split}
\?
where
\begin{subequations}\label{cCs}
\begin{gather}
\cC_{rr} := \left(\frac{3}{r} + \phi'\right)G_\GR(r) - \frac{3}{2}G_\GR'(r) -
h^{1/2}\bar{G}_\GR(r),\\
\begin{split}
\cC_{r\perp} &:= -\frac{\beta h^{1/2}}{r}G_\GR(r) + \frac{2 + r(\phi' + \psi')}{\beta h^{1/2}}G_\GR'(r)\\
&\,\quad - \frac{r}{\beta h^{1/2}}G_\GR''(r) + \frac{4 - 2r\phi'}{\beta}\bar{G}_\GR(r) - 2\frac{r}{\beta}\bar{G}'_\GR(r),
\end{split}\\
\begin{split}
\cC_\Lambda &:= \left(\frac{1}{\beta^2} - \frac{1}{2}\right)h^{1/2}G_\GR'(r)\\
&\,\quad + \frac{2r\phi'(3 - r\phi') + 2r\psi' - \beta^2h + 2}{\beta^2}\bar{G}_\GR(r)\\
&\,\quad - \frac{2r^2\phi'}{\beta^2}\bar{G}_\GR'(r),
\end{split}
\end{gather}
\end{subequations}
and we have integrated by parts twice to obtain the second equality in
Eq.~\eqref{Q22GR}, using the same argument as in our Newtonian calculation.

We now look at the maximum quadrupole. This is still given by the uniformly
maximally strained case: We have checked numerically that the
coefficients of the three
stress terms are always negative for all the background stars we consider. We thus
have a maximum quadrupole
given by inserting Eqs.~\eqref{ts}, \eqref{srr}, \eqref{srp}, and
\eqref{sLGR} into Eq.~\eqref{Q22GR}, yielding
\begin{widetext}
\<\label{Q22GR2}
\frac{|Q_{22}^\mathrm{max, GR}|}{\bar{\sigma}_\mathrm{max}} = \sqrt{\frac{32\pi}{15}}\int_0^R\mu(r)\biggl\{\left[\frac{6}{r}(h^{1/2} - 1) - 2\phi'\right]G_\GR(r) + \left[3 - \frac{r}{h^{1/2}}(\phi' + \psi')\right]G_\GR'(r)
+ \frac{r}{h^{1/2}}G_\GR''(r) + \cQ^\mathrm{stress}\biggr\}dr,
\?
where
\<
\cQ^\mathrm{stress} := 2\left[\frac{r\phi'(r\phi' - 3) - r\psi'  - 1}{h} + r\phi' + h^{1/2} + 1\right]\bar{G}_\GR(r) + 2r\left(\frac{r\phi'}{h} + 1\right)\bar{G}_\GR'(r)
\?
\end{widetext}
is the contribution from the stresses' own gravity. We have split it off both for
ease of notation and because it is negligible except for the
most massive and compact stars, as illustrated
below. The  contributions from the density and pressure perturbations are so much larger due to the factor of $1/\phi'$ present in $G_\GR$ [cf.\ Eqs.~\eqref{GGR} and~\eqref{GbGR}]. It is easy to see that Eq.~\eqref{Q22GR2} reduces to
Eq.~\eqref{Q22N2} in the Newtonian limit, where $h\to 1$,
and we can neglect the contributions involving $\phi'$, $\psi'$, and $\cQ^\mathrm{stress}$.

\begin{figure}[htb]
\begin{center}
\epsfig{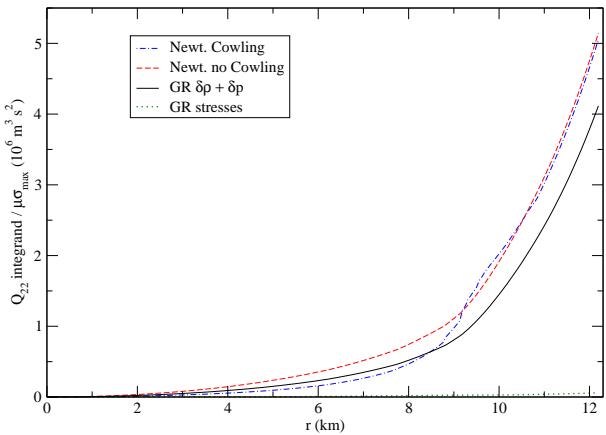}
\end{center}
\caption[$Q_{22}$ integrands for the SLy EOS and an
$0.500\Msolar$ star]{\label{Q22_integrands_SLy_2} The $Q_{22}$ integrands (without the factor of $\mu\bar{\sigma}_\mathrm{max}$) for the SLy EOS and an
$0.500\Msolar$ star with a compactness of $0.12$.}
\end{figure}

\begin{figure}[htb]
\begin{center}
\epsfig{file=Q22_integrands_SLy_h0_2e20.eps,width=8cm,clip=true}
\end{center}
\caption[$Q_{22}$ integrands for the SLy EOS and a
$1.40\Msolar$ star]{\label{Q22_integrands_SLy_4} The $Q_{22}$ integrands (without the factor of $\mu\bar{\sigma}_\mathrm{max}$) for the SLy EOS and a
$1.40\Msolar$ star with a compactness of $0.35$.}
\end{figure}

\begin{figure}[htb]
\begin{center}
\epsfig{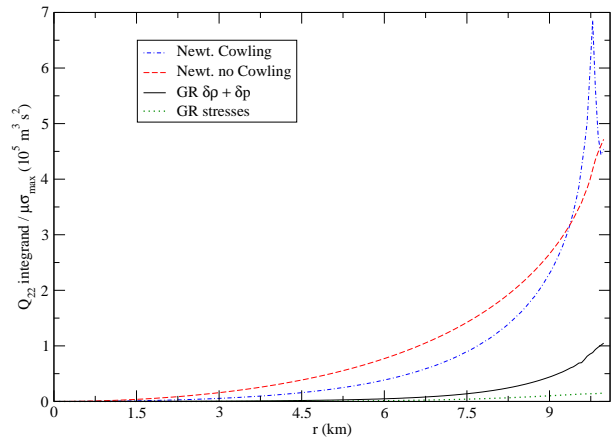}
\end{center}
\caption[$Q_{22}$ integrands for the SLy EOS and a maximum mass star]{\label{Q22_integrands_SLy_6} The $Q_{22}$ integrands (without the factor of $\mu\bar{\sigma}_\mathrm{max}$) for the SLy EOS and a maximum mass,
$2.05\Msolar$ star, with a compactness of $0.60$.}
\end{figure}

We now show how the relations between the different maximal-strain $Q_{22}$ Green functions
[given by the integrands in Eqs.~\eqref{Q22N2} and~\eqref{Q22GR2} without the factors of
$\mu$ (but with the overall prefactor)] vary with
EOS, as well as with the mass of the star for a given EOS.
This gives
an indication of how much difference the various approximations make in different situations. We start with the unified SLy EOS~\cite{DH}, obtained by Haensel
and Potekhin~\cite{HP}
(using the table provided by
the Ioffe group~\cite{HPY} at~\cite{Ioffe}), which is a standard choice for making predictions about crustal
quadrupoles (e.g., in Horowitz~\cite{Horowitz}, HJA, and our Sec.~\ref{results_crust}). Here we illustrate the
changes in the Green functions with mass for stars with masses ranging from $0.5\Msolar$ to the EOS's maximum mass of $2.05\Msolar$;
see Figs.~\ref{Q22_integrands_SLy_2},  \ref{Q22_integrands_SLy_4}, and \ref{Q22_integrands_SLy_6}. (All three Green functions agree extremely closely for stars around the EOS's minimum mass of $0.094\Msolar$, so we do not show this case, particularly because such
low-mass neutron stars are of unclear astrophysical relevance.)
These stars' compactnesses [defined in Eq.~\eqref{compactness}]
range from $0.12$ to
$0.6$. Note that Fig.~\ref{Q22_integrands_SLy_6} has a different vertical scale than the other two plots, due to the suppression of the quadrupole for massive, compact stars (discussed below). 

\begin{figure}[htb]
\begin{center}
\epsfig{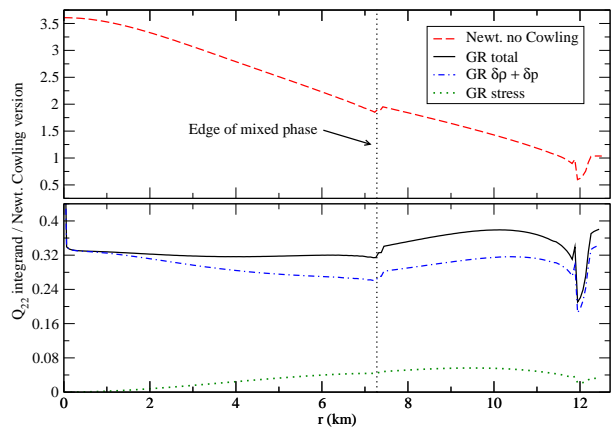}
\end{center}
\caption[Ratios of the $Q_{22}$ integrands for the Hy1 EOS and a maximum mass
star]{\label{Q22_integrands_Hy1_ratios} Ratios of the $Q_{22}$ integrands
with the Newtonian Cowling approximation integrand for the Hy1 EOS and a maximum mass,
$2.06\Msolar$ star, with a compactness of $0.49$. Note that the top and bottom plots have completely
separate vertical axis scalings.}
\end{figure}

We illustrate the ratios of the various $Q_{22}$ Green functions to the
Newtonian Cowling approximation one for the maximum mass ($2.05\Msolar$) hybrid
star using the Hy1 EOS (see Table~I in~\cite{J-MO1}) in Fig.~\ref{Q22_integrands_Hy1_ratios}.\footnote{As
discussed in~\cite{J-MO1}, for our low-density EOS, we use the same combination of
the Baym, Pethick, and Sutherland (BPS)~\cite{BPS} EOS for $n_B < 0.001\text{ fm}^{-3}$ and
the Negele and Vautherin~\cite{NV} EOS for $0.001\text{ fm}^{-3} < n_B < 0.08\text{ fm}^{-3}$
used by Lattimer and Prakash~\cite{LP2001} ($n_B$ is the baryon number density).
These were obtained from the table provided by
Kurkela~\emph{et al.}~\cite{Kurkelaetal} at~\cite{Kurkelaetal_URL}.
Bulk quantities of hybrid stars such as the mass and quadrupole moment (from core deformations) do not depend much on the precise choice of low-density EOS.} We see the overestimate of the Newtonian
no Cowling approximation calculation for perturbations in the core, particularly compared with the 
general relativistic (GR) version, and also see the
overestimate of the Newtonian Cowling approximation version for crustal perturbations. (We do not make
some sort of similar plot for the solid strange quark star case, since the expressions for the maximum
quadrupole in this case end up being rather different than the integrated-by-parts ones presented in the previous sections, as we shall see in Sec.~\ref{SQM_computation}.)

In all these cases, we compute the stellar background fully relativistically, using
the OV equations and identifying the OV equations' Schwarzschild
radial coordinate
with the Newtonian radial coordinate when necessary. We have used the enthalpy form of the OV
equations given by Lindblom~\cite{Lindblom} and implemented the inner boundary
condition by taking the star to have an inner core of radius $r_0 =
100\text{ cm}$, whose mass is given by $(4/3)\pi r_0^3\epsilon_0$, where
$\epsilon_0$ is the energy density corresponding to the central enthalpy that
parametrizes the solution. (The spike near the origin seen in the bottom plot in Fig.~\ref{Q22_integrands_Hy1_ratios} is due to this implementation of the inner boundary condition and has a
negligible effect on the computed maximum quadrupoles.) In all cases, we have used {\sc{Mathematica}}~7's
default methods to solve the differential equations, find roots, etc. We have computed as many derivatives
as possible analytically, to aid numerical accuracy, e.g., using the OV equations to substitute for derivatives of the pressure, and also using the Green function equations to
express second derivatives of the Green functions in terms of the functions themselves and their first derivatives.

\section{Results}
\label{results2}

\subsection{Maximum $Q_{22}$ for crustal deformations}
\label{results_crust}

Here we consider the maximum quadrupoles from elastic deformations of a
nonaccreted
crust in three possible situations, following HJA. In particular, we use the SLy EOS
(as do Horowitz~\cite{Horowitz} and HJA, though they do not refer to it by that name) and
impose two comparison crustal thicknesses to ascertain how much this
affects the maximum quadrupole. Here we use the
same rough
model for the crust's shear modulus used by HJA. We
also consider the more detailed model for the shear modulus
obtained using the crustal composition provided by Douchin and
Haensel~\cite{DH} (also used by Horowitz~\cite{Horowitz} and HJA).
Here the crust's thickness is fixed to the value given in that work. In this case,
we also consider a different high-density EOS that yields much less compact stars with larger crusts.

\begin{figure}[htb]
\begin{center}
\epsfig{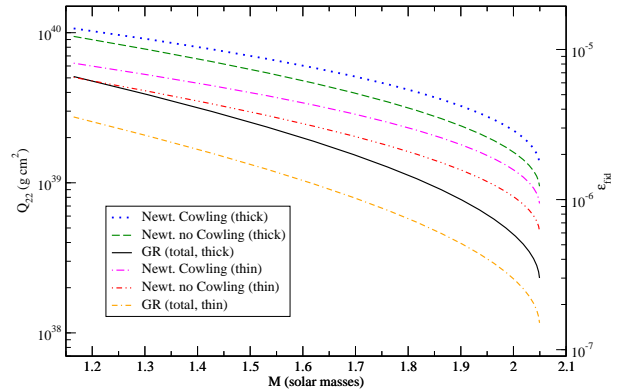}
\end{center}
\caption{\label{Q22s_vs_M_SLy} The Newtonian Cowling, Newtonian no
Cowling, and full relativistic (including stress contributions) values for the
maximum quadrupole deformations (and fiducial ellipticity) due to crustal
stresses versus mass for two choices of crustal thickness. These are computed using the SLy EOS with the rough HJA recipe for the shear
modulus and a breaking strain of $0.1$.}
\end{figure}

Specifically, the two comparison crustal thicknesses are given by taking the base of the
crust to occur at densities of $2.1\times10^{14} \text{ g cm}^{-3}$ (thick
crust, for comparison with UCB) or $1.6\times10^{14} \text{ g cm}^{-3}$ (thin
crust, following a suggestion by Haensel~\cite{Haensel}), while Douchin and
Haensel place the bottom of the crust at a density of
$1.28\times10^{14} \text{ g cm}^{-3}$.
For the two comparison cases, we take the shear modulus to be
$10^{16}\text{ cm}^2\text{ s}^{-2}$ times the star's density (in $\text{g cm}^{-3}$). As
illustrated in HJA's Fig.~2, this is an underestimate of $<50\%$, except at the very extremes of the density range considered.\footnote{Note that Fig.~3 in HJA is not in agreement with their Fig.~2. When we reproduce those figures, we
find that the ratio $\mu/\rho$ is considerably closer to $10^{16}\text{ cm}^2\text{ s}^{-2}$ over all the density range than the trace shown in HJA's Fig.~3, so their approximation is better than it would appear from that figure.}
We plot the quadrupole moment and ellipticity for these two cases for masses between
$\sim 1.2\Msolar$ (around the minimum observed neutron star mass---see~\cite{Lattimer_table})
and the SLy EOS's maximum mass of $2.05\Msolar$ in Fig.~\ref{Q22s_vs_M_SLy}.

In addition to the quadrupole moments, we also show the fiducial ellipticity
$\epsilon_\mathrm{fid} = \sqrt{8\pi/15}Q_{22}/I_{zz}$
[e.g., Eq.~(2) of~\cite{OwenPRL}~].
Here  $I_{zz}$ is the star's principal moment of
inertia, for which we use the fiducial value of $I_{zz} = 10^{38} \text{ kg m}^2 =
10^{45} \text{ g cm}^2$ used in the LIGO/Virgo papers rather than the true value for a given mass and EOS,
which can be greater by a factor of a few.
We do this for easy comparison with the observational
papers, since they frequently quote
results in terms of this fiducial ellipticity instead of the quadrupole
moment, which is the quantity truly measured.

\emph{Nota bene} (N.B.):\ We present these fiducial ellipticities \emph{only} for comparison with LIGO/Virgo
results, not to give
any indication of the size of the deformation. While the true ellipticity
gives a measure of the size of the deformation in the Newtonian case (up to ambiguities from the fact that the true density distribution is nonuniform), it does not do so in
any obvious way in the relativistic case. Nevertheless, the relativistic shape of the star's surface can be obtained from its quadrupole deformation, as shown in~\cite{NKJ-M_shape}. However, if one wished to know, for instance, how much the star is deformed as a function of radius, one would need to calculate this using a detailed relativistic theory of elasticity to relate the stresses to the matter displacements, as in Penner~\emph{et al.}~\cite{Penneretal}.

In the more detailed
case, we use the HJA version of the Ogata and Ichimaru~\cite{OI} shear modulus,
combined with the Douchin and Haensel~\cite{DH} results for the crust's
composition.
This is [correcting a typo in HJA's Eq.~(20)],
\<\label{mueff}
\mu_\eff = 0.1194\left(\frac{4\pi}{3}\right)^{1/3}\left(\frac{1-X_n}{A}n_b\right)^{4/3}(Ze)^2,
\?
where $X_n$ is the fraction of neutrons outside of nuclei, $A$ and $Z$ are the
atomic and proton number of the nuclei, respectively, $n_b$ is the baryon number density, and
$e$ is the fundamental charge.

Since HJA's study, there have been a few improvements
to the Ogata and Ichimaru result: Horowitz and Hughto~\cite{HH} have computed
the effects of charge screening, finding a $\sim 7\%$ reduction in the shear
modulus. Baiko~\cite{BaikoCPP} has also considered a relativistic model
for the electron polarizability and arrived at similar conclusions. Indeed, Baiko's results suggest that screening will yield an even smaller
correction in the innermost portion of the crust, where the shear modulus is the largest, and the
electrons are the most relativistic, with a relativity parameter over an order of magnitude larger than the largest
Baiko considers. (However, the ion charge numbers are also almost always somewhat greater than the largest Baiko considers,
particularly at the very innermost portion of the crust, which will tend to increase the effect.) 

Baiko~\cite{Baiko} has also recently computed quantum corrections, and finds
that they reduce the shear modulus by up to $\sim 18\%$ in some regimes.
However, in our case, the reduction will be much smaller, based on the scaling
of $\rho^{1/6}/(ZA^{2/3})$ given near the end of Baiko's Sec.~6. Even though
our densities are over an order of magnitude greater, the nuclei we
consider are also over an order of magnitude more massive than the ${}^{12}$C
composition Baiko considers, so
the quantum mechanical effects end up being reduced by about an order of
magnitude from the number Baiko quotes.
We thus use the same Ogata and Ichimaru result used by HJA, noting that the
resulting quadrupoles might be reduced by less than $10\%$ due to charge screening
and quantum effects---an error which is small compared to other
uncertainties, such as crust thickness and the composition of dense matter. Indeed, there
is a factor of $\sim 2$ uncertainty in the shear modulus due to angle averaging (even
disregarding whether the implicit assumption of a polycrystalline structure for the crust is
warranted): As shown by
Hill~\cite{Hill}, the Voigt average used by Ogata and Ichimaru is an upper bound on the
true shear modulus of a polycrystal. A lower bound is given by the Reuss average (also
discussed in Hill~\cite{Hill}), for which the prefactor in Eq.~\ref{mueff} would be
$0.05106$. 

Note that there would be even
further corrections to the shear modulus due to pasta phases (see~\cite{PP}), but such phases are not present in the Douchin and Haensel model~\cite{DH}. We also note that the Douchin and Haensel results only include the very
innermost portion of the outer crust. However, this lack of coverage has a negligible effect on the final
results for the quadrupoles, since the neglected region has at most half the radial extent of the inner crust and the shear modulus in this region is orders of magnitude below its maximum value at the bottom of the inner crust. We have checked this explicitly using the detailed calculations
of the outer crust composition due to R{\"u}ster, Hempel, and Schaffner-Bielich~\cite{RHS-B}, available at~\cite{HempelURL}.

\begin{figure}[htb]
\begin{center}
\epsfig{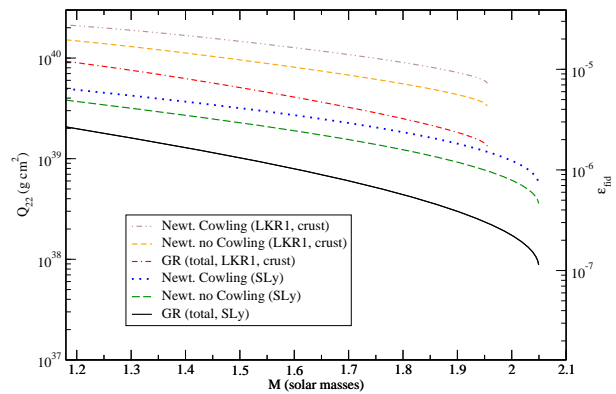}
\end{center}
\caption{\label{Q22s_vs_M_SLy_DH_HJA} The Newtonian Cowling, Newtonian no
Cowling, and full relativistic (including stress contributions) values for the
maximum quadrupole deformations (and fiducial ellipticity) due to crustal
stresses versus mass, for the SLy EOS with the detailed Douchin and Haensel + Ogata and
Ichimaru model for the shear modulus and a breaking strain of $0.1$, plus the crustal quadrupoles for the LKR$1$ EOS with the same crustal model.
}
\end{figure}

We plot the maximum quadrupole and ellipticity in the three approximations for the detailed shear modulus
model in Fig.~\ref{Q22s_vs_M_SLy_DH_HJA}. Here we show these for the SLy EOS proper, and also for
a high-density EOS that yields much less compact stars (and a crust that is $\sim 2$ times as thick), and thus larger maximum quadrupoles. For the latter EOS, we have chosen (for simplicity) the LKR1 hybrid EOS from~\cite{J-MO1}---the maximum compactnesses for the two EOSs are $0.60$ (SLy) and $0.43$ (LKR$1$). (We show the much larger quadrupoles that could be
supported by the mixed phase in the core for the LKR$1$ EOS in Fig.~\ref{Q22_vs_M_EOSs}, but here just show the crustal quadrupoles using the Douchin and Haensel model for the crust.)

In all of these crustal results, in addition to the expected relativistic suppression of the quadrupole (which
becomes quite dramatic for compact, high-mass stars), we also find that the
Newtonian Cowling approximation slightly overestimates the quadrupole (by
$\sim 25$--$50\%$), as observed by HJA (though they found the overestimate to be
considerably greater, around a factor of at least a few). This overestimate is due to the cancellation of
contributions from $\rho'$ when one drops
the Cowling approximation (see the discussion at the end of Sec.~\ref{Newt}).
The overall decrease in the maximum crustal quadrupole with mass is due primarily to the fact that the crust thins by a factor of $\sim 4$ (SLy) or $\sim 2$ (LKR1) in going from a $1\Msolar$ star to
the maximum mass star, though the quadrupole itself receives even further suppressions with mass due to relativistic
effects and an increased gravitational field.

\subsection{Maximum $Q_{22}$ for hybrid stars}
\label{results_hybrid}

\begin{figure}[htb]
\begin{center}
\epsfig{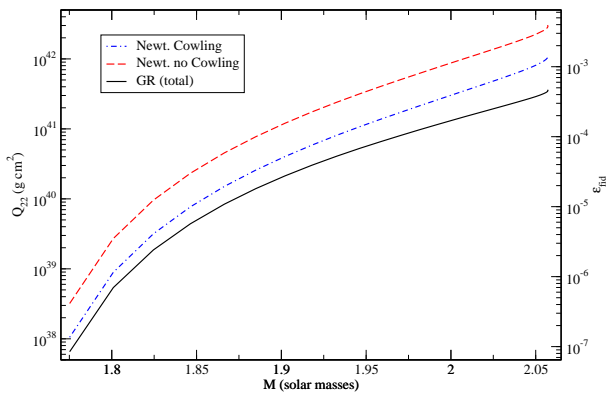}
\end{center}
\caption{\label{Q22s_vs_M_Hy1_sigma80} The Newtonian Cowling, Newtonian no
Cowling, and full relativistic (including stress contributions) values for the
maximum quadrupole deformations (and fiducial ellipticity) of hybrid stars versus
mass, using the Hy1 EOS with a surface tension of
$\sigma = 80\text{ MeV fm}^{-2}$ and a breaking strain of $0.1$.}
\end{figure}

Here we display the maximum quadrupole deformations as a
function of stellar mass for each of the hybrid EOS parameter sets considered
in~\cite{J-MO1}. (N.B.: Most of the results from~\cite{J-MO1} we use or refer to here were
corrected in the erratum to that paper.) We start by showing these values calculated in the various
approximations using the Hy1 EOS (with a surface tension of
$\sigma = 80\text{ MeV fm}^{-2}$; see Table~I in~\cite{J-MO1}) in
Fig.~\ref{Q22s_vs_M_Hy1_sigma80}, and then
restrict our attention to the relativistic results. (The relation between the results of the different approximations is roughly
the same for all the hybrid EOSs we consider.)
Here the maximum quadrupoles increase with mass, since the volume of mixed phase increases with mass, and this is more than enough to offset the suppressions due to relativity and the increased gravitational field.

\begin{figure}[htb]
\begin{center}
\epsfig{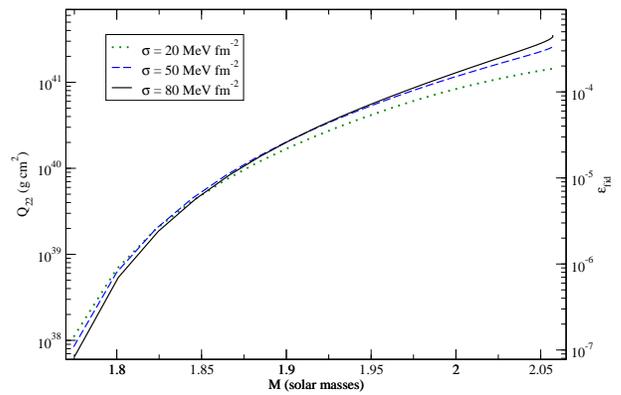}
\end{center}
\caption{\label{Q22_vs_M_Hy1_sigmas} The full relativistic maximum quadrupole deformations (and fiducial ellipticity) of hybrid
stars versus mass, using the Hy1 EOS with various surface tensions $\sigma$
and a breaking strain of $0.1$.}
\end{figure}

We also show how the maximum relativistic quadrupole
varies with the surface tension for the Hy1 EOS in
Fig.~\ref{Q22_vs_M_Hy1_sigmas}.  The slightly larger quadrupoles for lower surface tensions
at low masses are expected, due to a slightly larger shear modulus at low
pressures for lower surface tensions---see Fig.~10 in~\cite{J-MO1}. In fact, despite differences of
close to an order of magnitude in the high-pressure shear modulus for the Hy$1$ EOS
in going from a surface tension of $20\text{ MeV fm}^{-2}$ to one of
$80\text{ MeV fm}^{-2}$ (see Fig.~10 in~\cite{J-MO1}), the
differences in the resulting maximum quadrupoles are at most
a factor of a few (for large masses). This is not unexpected: These quantities
are dominated by the portions of the mixed phase further out in the star, where
the shear moduli have a much weaker dependence on the surface tension.
(Additionally, the fact that larger surface tensions lead to smaller shear moduli at low
pressures helps to minimize the effect, though the maximum
quadrupoles still increase with increasing surface tension for high masses, as
expected.) 

\begin{figure}[htb]
\begin{center}
\epsfig{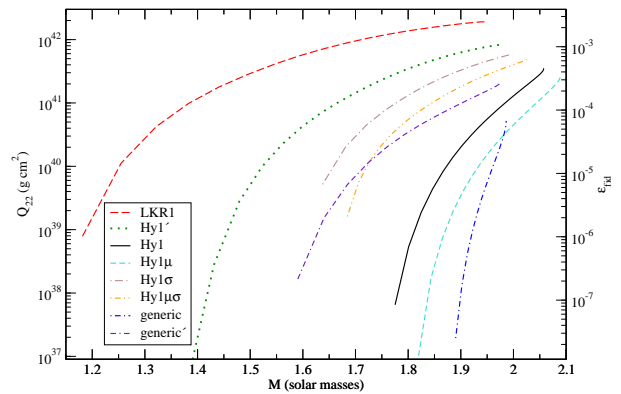}
\end{center}
\caption{\label{Q22_vs_M_EOSs} The full relativistic maximum quadrupole deformations (and fiducial ellipticity) of hybrid
stars versus mass, using the EOSs from Table~I in~\cite{J-MO1}, all with a surface tension of
$\sigma = 80\text{ MeV fm}^{-2}$
and a breaking strain of $0.1$.}
\end{figure}

Finally, we show the maximum quadrupoles for different
hybrid EOSs in Fig.~\ref{Q22_vs_M_EOSs}. (Note that these curves start somewhat above the minimum
masses for which the mixed phase is present, since we are mostly interested in the significantly larger
maximum quadrupoles possible for larger masses.) The
considerable differences are due primarily to the substantial variations in the extent
of the mixed phase in stable stars with EOS parameters as well as the
EOS dependence of the stars'
compactnesses (see Table~I in~\cite{J-MO1}), not to variations in the magnitude of the shear modulus for a
given quark matter fraction (compared in Fig.~12 in~\cite{J-MO1}). In particular, the LKR1 EOS produces stars with a very large region of mixed
phase---up to $72.5\%$ of the star's radius---and a (relatively) small maximum
compactness---only $0.433$. (Note that our quadrupole curve for
the LKR1 EOS ends slightly short of the EOS's maximum mass of $1.955\Msolar$,
only going to $1.948\Msolar$, due to problems with the numerics.)

N.B.:\ These maximum quadrupoles may all be overly optimistic. First, as was discussed in Sec.~\ref{results_crust}, the averaging used to obtain the effective shear modulus only gives an upper bound on the true shear modulus of a
polycrystal. (We do not quote results for the Reuss lower
bound here, since it is only straightforward to obtain for the three-dimensional droplet phases. However, 
we shall note that preliminary investigations, using the Reuss bound for the droplet phases, and the
Voigt bound for the rest, give reductions in the maximum quadrupoles of up to $\sim 5$ for lower masses.)

Second, the relatively large value we have chosen for the surface tension also increases the maximum
quadrupoles, while recent calculations place the surface tension on the low side ($\sim 10$--$30\text{ MeV fm}^{-2}$)---see~\cite{PKR} for the latest results. Nevertheless, as we show in the Appendix, the mixed phase is nevertheless favored by global energy
arguments even for these large surface tensions. The maximum quadrupoles are also affected by the
method of EOS interpolation and the lattice contributions to the EOS, as is illustrated in the Appendix, though
the largest change is only $\sim40\%$ (at least for the LKR$1$ and Hy$1'$ EOSs, the two EOSs that yield the largest quadrupoles).

Note that LIGO's current upper limits on fiducial ellipticity 
in the most interesting cases (the Crab pulsar, PSR~J0537--6910, and
Cas~A)~\cite{LIGO_psrs2010, LIGO_CasA} are $\sim 10^{-4}$, corresponding to a
quadrupole moment of $\sim 10^{41} \text{ g cm}^2$.
The first hybrid star estimate by \citet{OwenPRL} was an order of magnitude
lower.
Thus our new results here show that current LIGO upper limits are interesting
not only for quark stars but also for hybrid stars, at least high-mass ones.
Indeed, the most extreme case we consider, the LKR$1$ EOS with high surface tensions, gives maximum
quadrupoles of a $\text{few} \times 10^{42}\text{ g cm}^2$, which are above and therefore relevant to the
limits set by Virgo for the Vela pulsar~\cite{LIGO_Vela}. 

\subsection{Maximum $Q_{22}$ for crystalline color superconducting quark stars}
\label{SQM_computation}

Here we consider stars made of crystalline color superconducting quark matter, for which the
shear modulus has been estimated by Mannarelli, Rajagopal, and Sharma~\cite{MRS}.\footnote{This
estimate is not angle averaged, but Mannarelli, Rajagopal, and Sharma's calculation has relatively large uncontrolled remainders, so we do not worry about the effects of angle averaging here.}
[See Eq.~(1) in Haskell~\emph{et al.}~\cite{Haskelletal} for the expression in cgs units.] Such
stars have also been treated (with varying degrees of sophistication) by Haskell~\emph{et al.}~\cite{Haskelletal}, \citet{Lin}, and~\citet{KS}. However, only Lin considers the case of a solid quark star, as we will do here, and does so using quite a rough model. (The others consider crystalline color superconducting
cores in hybrid stars.)

Since strange quark stars have a nonzero surface density---and solid quark stars have a nonzero
surface shear modulus, with the standard density-independent treatment of the superconducting
gap---we have to make some changes to our previously obtained expressions
in order to treat them.

First, the outer boundary condition changes. The potential (in the Newtonian case) and metric
perturbation (in the GR case) are no longer continuous at the star's surface, due to the presence
of $\rho'$ in both equations [see Eqs.~\eqref{Leq} and~\eqref{H0_S-L}]. As discussed in Hinderer~\emph{et al.}~\cite{Hindereretal} (following Damour and Nagar~\cite{DN}), one can obtain the distributional contribution to the boundary conditions [Eqs.~\eqref{Newt_BC} and~\eqref{H0_BC}] using the usual procedure of integrating the defining
differential equation over $[R-\epsilon,R+\epsilon]$ and taking the limit $\epsilon\searrow 0$.
In the Newtonian case, this gives [defining $\rho_-$ as the density immediately inside the star's
surface and $R^-$ to mean evaluation at $R-\epsilon$ in the limit $\epsilon\searrow 0$]
\<
\delta\Phi'(R^-) = \left[\frac{4\pi G}{g(R)}\rho_- - \frac{3}{R}\right]\delta\Phi(R),
\?
and in the GR case, we have (with $G = 1$)
\<
H_0'(R^-) = H_{0,\mathrm{old}}'(R) + \frac{4\pi h}{\phi'(R)}\rho_-H_0(R),
\?
where $H_{0,\mathrm{old}}'(R) $ is computed using Eq.~\eqref{H0_BC}.
We thus make the replacement $3RF(R) \to [3 - 4\pi G \rho_- R/g(R)]RF(R)$ in the expression 
for the Newtonian Green function [Eq.~\eqref{cG_F}], and the replacement
$H_0'(R) \to H_{0,\mathrm{old}}'(R) + 4\pi h\rho_-H_0(R)/\phi'(R)$ in the GR case [Eq.~\eqref{cG_GR}].
These changes in the boundary conditions increase the maximum quadrupole by a factor of $\lesssim 2$ in the example case considered below; the largest effect is for the least massive stars considered.

Second, we would have to keep the boundary terms at the outer boundary when integrating by parts to
obtain the expressions for the maximum quadrupole, since the shear modulus no longer vanishes at the
star's surface. However, since here the shear modulus is smooth, it is numerically preferable not to perform any integration by parts, thus avoiding potential problems with large cancellations between the surface and integrated terms. In this case, the expressions for the quadrupole assuming the UCB maximum
uniform strain are [cf.\ Eqs.~\eqref{Q22N2} and \eqref{Q22GR2}]
\<
\label{Q22UCBs}
\frac{|Q_{22}^{\text{UCB strain}, N}|}{\bar{\sigma}_\mathrm{max}} = \sqrt{\frac{32\pi}{15}}\int_0^RG_N(r)[r\mu''(r) - \mu'(r)]dr
\?
and
\<
\begin{split}
\frac{|Q_{22}^\text{UCB strain, GR}|}{\bar{\sigma}_\mathrm{max}} &= \sqrt{\frac{32\pi}{15}}\int_0^R\bigl[G_\GR(r)\cI_{[\delta\rho,\delta p]}^\mathrm{UCB}(r)\\
&\quad + \bar{G}_\GR(r)\cI_{[t]}^\mathrm{UCB}(r)\bigr]dr,
\end{split}
\?
where
\begin{widetext}
\begin{subequations}
\begin{align}
\cI_{[\delta\rho,\delta p]}^\mathrm{UCB}(r) &:= \left[\frac{6}{r}(h^{1/2} - 1) - 2\phi' + \frac{r\phi''  + \phi'(1 - r\psi') - \psi'}{h^{1/2}}\right]\mu(r) + \left[\frac{2 + r(\phi' - \psi')}{h^{1/2}} - 3\right]\mu'(r) + \frac{r\mu''(r)}{h^{1/2}},\\
\cI_{[t]}^\mathrm{UCB}(r) &:= 2\left\{\frac{r\phi'[r(\phi' + 2\psi') - 5] - r\psi' - r^2\phi'' - 1}{h} + r\phi' + h^{1/2}\right\}\mu(r) - 2r\left(\frac{r\phi'}{h} + 1\right)\mu'(r).
\end{align}
\end{subequations}
\end{widetext}

However, these expressions will not actually yield the maximum quadrupole in this case, due to an important
difference between the cases where the shear modulus vanishes at the star's surface and those where it does not. It is simplest to see this in the
Newtonian case for a star with a constant shear modulus: Since the UCB maximum strain expression~\eqref{Q22UCBs} only
depends upon derivatives of the shear modulus, it
predicts a \emph{zero} maximum quadrupole, which seems absurd. One can, however, make a small
adjustment to the form of the maximum strain one considers to yield a nonzero quadrupole in this case. This modification will also yield considerably larger maxima in the realistic case we
consider, as well, where the shear modulus is close to constant---it decreases by less than a factor of $2$ in going from
the star's center to its surface in the example case we consider below.

Specifically, in the case of a slowly varying shear modulus, with $\mu(r)\gg|r\mu'(r)|,|r^2\mu''(r)|$, appropriate for strange quark stars, we want the terms involving $\mu$ itself to be
largest. The appropriate choice for the strain in this case is most readily apparent
from inspection of the Newtonian expression for the maximum quadrupole in terms of the stress tensor components, Eq.~\eqref{Q22N1}. We want the maximum contribution from the undifferentiated terms, which implies that we want $t_{rr}$ and $-t_{r\perp}$ to be as large as possible. For $t_\Lambda$, we note that since $\mu'(r)<0$,
we also want $-t_\Lambda$ to be as large as possible. Realizing that we can freely change the sign of any of
the $\sigma_\bullet$ that give maximum uniform strain [given for the Newtonian case in Eqs.~\eqref{sNewt}; cf.\ Eq.~(65) in UCB], we thus reverse the sign of $\sigma_{r\perp}$ and $\sigma_\Lambda$. [The same logic holds for the more involved GR case, as well, where
the appropriate expression for $\sigma_\Lambda$ will be the negative of Eq.~\eqref{sLGR}.]

The resulting expressions for the putative maximum quadrupole in this case are thus
\<\begin{split}
\frac{|Q_{22}^{\text{mod.\ strain}, N}|}{\bar{\sigma}_\mathrm{max}} &= \sqrt{\frac{32\pi}{15}}\int_0^RG_N(r)\biggl[\frac{12}{r}\mu(r) + 5\mu'(r)\\
&\quad +r\mu''(r)\biggr]dr
\end{split}
\?
and
\<
\begin{split}
\frac{|Q_{22}^{\text{mod.\ strain, GR}}|}{\bar{\sigma}_\mathrm{max}} &=  \sqrt{\frac{32\pi}{15}}\int_0^R\bigl[G_\GR(r)\cI_{[\delta\rho,\delta p]}^\mathrm{mod}(r)\\
&\quad + \bar{G}_\GR(r)\cI_{[t]}^\mathrm{mod}(r)\bigr]dr,\end{split}
\?
where
\begin{widetext}
\begin{subequations}
\begin{align}
\cI_{[\delta\rho,\delta p]}^\mathrm{mod}(r) &:= \left[\frac{6}{r}(h^{1/2} + 1) + 2\phi' + \frac{r\phi'' + \phi'(1 - r\psi') - \psi'}{h^{1/2}}\right]\mu(r) + \left[\frac{2 + r(\phi' - \psi')}{h^{1/2}} + 3\right]\mu'(r) + \frac{r\mu''(r)}{h^{1/2}},\\
\cI_{[t]}^\mathrm{mod}(r) &:= - 2\left\{\frac{ r\phi'[5- r(\phi'+2\psi')] + r\psi' + r^2\phi'' + 1}{h} - r\phi' + h^{1/2}\right\}\mu(r)
- 2r\left(\frac{r\phi'}{h} + 1\right)\mu'(r).
\end{align}
\end{subequations}

\end{widetext}
In principle, these merely give a lower bound on the
maximum quadrupole, unlike the case in which the shear modulus vanishes below the surface, where there is a firm argument that maximum uniform strain maximizes the quadrupole. However, even if they do not give the absolute
maximum, they should be quite close for cases like the one we consider here, where the shear modulus varies quite slowly.

\begin{figure}[htb]
\begin{center}
\epsfig{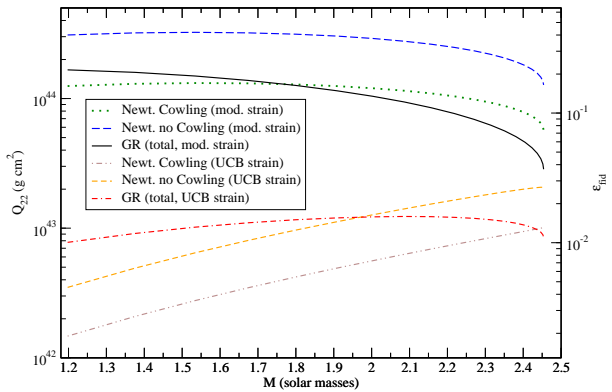}
\end{center}
\caption{\label{Q22s_vs_M_SQM1} The Newtonian Cowling, Newtonian no
Cowling, and full relativistic (including stress contributions) values for the
quadrupole deformations (and fiducial ellipticity) of maximally strained strange quark stars versus
mass, using the EOS discussed in the text with a breaking strain of $0.1$. We show these both for the
standard UCB uniform maximum strain, and our modification that yields significantly larger quadrupoles in this case.}
\end{figure}

Applying these expressions to a specific case, we use the strange quark matter EOS calculated by Kurkela,
Romatschke, and Vuorinen (KRV)~\cite{KRV}, generating an EOS for the parameter values of
interest using the {\sc{Mathematica}} notebooks available at~\cite{KRV_URL}.
The relevant parameters are the values of the $\overline{\mathrm{MS}}$ renormalization point,
$\Lambda_{\overline{\mathrm{MS}}}$, and the strange quark mass, $m_s$, both at a scale of $2$~GeV, along with the coefficient in the relation between the renormalization scale and the
quark chemical potential, $X$, the color superconductivity gap parameter, $\Delta$ (taken to be
independent of density),\footnote{Note that $\Delta$ enters the KRV EOS through a color flavor locked (CFL)
pressure term. This is not quite appropriate for the crystalline color superconducting phase we consider here, since it assumes that all the quarks pair, while only some of them pair in the crystalline phase. However, as
discussed in Sec.~VI~B of~\cite{Alford2008}, the condensation energy of the crystalline phases is easily
$1/3$ to $1/2$ that of the CFL phase with zero strange quark mass, which is the pressure
contribution used by KRV. We have thus not altered this term in our calculations, since the contribution is
already approximate, in that it assumes a density-independent gap parameter.
Moreover, we only consider a fairly low value of $\Delta$,
while Knippel and Sedrakian~\cite{KS} suggest that the crystalline phase might be favored up to $\Delta = 100$~MeV. Our EOS may thus simply correspond to a slightly larger value of $\Delta$, which would 
increase the maximum quadrupole, since the shear modulus scales as $\Delta^2$.} and the
minimal quark chemical potential at which strange quark matter exists, $\mu_{q,\mathrm{min}}$.
We consider the EOS obtained by choosing $\Lambda_{\overline{\mathrm{MS}}} = 355$~MeV, $m_s = 70$~MeV, $X=4$, $\Delta = 10$~MeV, and $\mu_{q,\mathrm{min}} = 280$~MeV.
This parameter set yields a maximum mass of $2.45\Msolar$, with a maximum compactness of $0.467$. 

These parameter choices were generally inspired by those considered at~\cite{KRV_URL}, though with a smaller value of $\Delta$, to place us well within the crystalline
superconducting regime. However, as Knippel and Sedrakian~\cite{KS} suggest, the crystalline phase
could still be favored for considerably larger $\Delta$s, up to $\sim 100$~MeV, for the low-temperature
case relevant for neutron stars. We thus note that increasing $\Delta$ decreases the maximum mass, and increases the
maximum quadrupole, though the latter is increased by considerably less than the na{\"\i}ve scaling of $\Delta^2$ one would
expect from the scaling of the shear modulus, likely due to the increased compactness of the stars with
larger $\Delta$s: For $\Delta = 100$~MeV, we have a maximum mass and compactness of $2.12\Msolar$ and $0.508$, respectively, and a maximum
quadrupole of $\sim 3.5\times 10^{45}\text{ g cm}^2$ for a $1.4\Msolar$ star, $\sim 20$ times that for $\Delta = 10$~MeV. However, one must bear in
mind that our perturbative treatment starts to become questionable with such large gap parameters, for which the maximum shear stresses are more than $10\%$ of the background's energy density. The uncontrolled remainders in the Mannarelli,
Rajagopal, and Sharma~\cite{MRS} calculation of the shear modulus also increase as the gap parameter
increases.

We show the quadrupole for a maximally uniformly strained star in the three approximations (Newtonian
Cowling, Newtonian no Cowling, and GR) for both the UCB and modified maximum strain choices for this EOS in Fig.~\ref{Q22s_vs_M_SQM1}. Here we have used a breaking strain of $0.1$, by the same high
pressure argument as in the mixed phase case. (While the very outermost portions of the star are at low
pressure, the parts that are at a lower pressure than the crustal case for which the $0.1$ breaking strain
was calculated make negligible
contributions to the quadrupole.)

N.B.:\ To obtain the EOS used for this figure, we made some slight modifications to the KRV {{\sc{EoScalc}}} {\sc{Mathematica}} notebook so that it would output particle number densities on a denser mesh for low strange quark chemical potentials. This then gave an EOS table with better low-pressure coverage than their default settings produced. We still needed to perform
an extrapolation of the EOS to zero pressure, where we found that a linear extrapolation of the energy
density and quark chemical potential in terms of the pressure using the lowest two entries of the table provided a good fit.
(More involved approaches involving fitting to more points and/or a quadratic extrapolation produce very similar results.)

Additionally, it is worth pointing out that the applying the KRV results to compact stars
pushes their second-order perturbative calculation towards the edge of its domain of validity. However,
in our case, the smallest value of the quantum chromodynamics (QCD) renormalization scale we consider is
$1.12$~GeV, at which value the QCD coupling constant is $\sim 0.45$. Thus, the uncontrolled remainders in
the expansion are suppressed by at least a factor of $\sim 0.1$. (While Rajagopal and Shuster~\cite{RS} find that perturbative QCD calculations of the color superconducting gap are only reliable at energy scales of
$\gtrsim 10^5$~GeV, the specifics of this calculation are rather different from the calculation of the EOS we
are considering here, where the gap is taken as an input parameter.) While it is unreasonable to expect
this calculation to be a truly accurate description of strange quark matter, it is not clear that any of the alternative descriptions of strange quark matter are \emph{a priori} guaranteed to be a better description of the physics, given the very considerable uncertainties associated
with this phase of matter.

\section{Discussion}
\label{discussion}

Previous studies of the tidal and magnetic deformations of compact
stars have found similar relativistic suppressions of quadrupole moments with
compactness.
In the tidal case, see the Love number computations in~\cite{Hinderer, BP, DN, Hindereretal, PPL}.
In the case of magnetic deformations, the expected suppressions are seen in,
e.g., \cite{IS, CFG, YKS, FR}.
In fact, since the largest compactness considered in these latter papers is only
$0.48$ (in~\cite{IS}), one imagines that they overestimate the maximum quadrupoles
by at least a factor of a few for more compact stars (for a fixed magnitude of magnetic
field).

As was argued by Damour and Nagar~\cite{DN} in the tidal case, all these suppressions are primarily related to the ``no-hair'' property of black holes: The largest relativistic suppression we find comes from the boundary conditions
[through the $H_0(R)$ and $H_0'(R)$ in the Green function's denominator---see Eq.~\eqref{cG_GR}], where
one matches on to the external vacuum spacetime. For instance, for the SLy EOS's maximum
compactness of $0.6$, $H_0(R)$ and $H_0'(R)$ are $\sim 3.5$ and $\sim 6$ times their
Newtonian values
[which can be obtained from the first term of Eq.~(21) in Hinderer~\cite{Hinderer}~].
In fact, these ratios go to infinity in the formal black hole limit, where the compactness
approaches unity, as required by the
no-hair property, and discussed by Damour and Nagar~\cite{DN} (see
their Secs.~IV~C and VII~A, but note that their definition for the compactness is half of ours). This implies that the stiffness of spherically symmetric
curved vacuum spacetime suppresses the quadrupole. The quadrupole is also
suppressed by a larger effective gravitational acceleration (given by $\phi'$), which appears
in the denominator of $G_\GR$, replacing the Newtonian $g(r)$ [cf.\ Eqs.~\eqref{GN} and~\eqref{GGR}]. (But
recall that we always compute the background stellar structure relativistically, so this
larger acceleration \emph{only} affects the perturbation equations, and not, e.g., the thickness of
the crust for a given mass and EOS, which is the same in both the Newtonian and
relativistic calculations of the quadrupole.) 

Our results imply that nearly all of the Newtonian computations of quadrupoles due to elastic
deformations of relativistic stars overestimate the quadrupole moment, often by at
least a factor of a few. The only exceptions we have found are
for low-to-mid mass strange quark stars and for elastic stresses in the cores of neutron stars around $0.5\Msolar$. In
both of these cases, the Newtonian Cowling
approximation is a slight underestimate for contributions to the quadrupole, though the Newtonian no Cowling
version is still an overestimate. See Fig.~\ref{Q22_integrands_SLy_2} for an illustration in the core case;
but note that neutron stars with such low masses are not known to exist in nature.
The overestimate from performing a Newtonian Cowling approximation calculation 
can be $\sim 6$ for massive stars whose quadrupole is
being generated by an elastic deformation near the
crust-core interface, as considered by UCB and others. This is due in part to the
sudden changes in density at that interface entering directly through $g''$, as
discussed at the end of Sec.~\ref{Newt}.

However, the calculations by Horowitz~\cite{Horowitz} for crustal deformations of very low mass stars only receive negligible
corrections (of $\lesssim5\%$), since he considers
compactnesses of $\sim 0.01$. In fact, one makes even
smaller errors in using the Cowling approximation to treat these stars,
since the changes in density in the crust (times $4\pi G r^2$) are much smaller than the star's gravitational field there.

No neutron stars with such low masses have ever been observed (nor is there
a compelling mechanism for forming them). Nevertheless, Horowitz remarks that
gravitational wave detection of
gravitational waves from elastically deformed neutron stars will,
\emph{ceteris paribus}, be biased towards low(er) mass neutron stars, if one
considers deformations generated by crustal stresses. This is an important 
point, particularly when considering the astronomical interpretation of
detections (or even upper limits), and the results we present here make the
bias against high-mass stars even stronger. (This bias also applies to solid
quark stars, though there it is rather weak. It does \emph{not} apply to hybrid
stars, however, where it is high-mass stars that can sustain the largest quadrupoles.)

Of course, one must remember that all of these values are maxima, assuming a
maximally strained star, while there is no reason, \emph{a priori}, for a given star to be
maximally strained.
Moreover, as UCB and HJA note, these calculations assume
that all the strain goes into the $l = m = 2$ perturbation, though strain in
other modes (e.g., the $l = 2$, $m = 0$ mode due to rotation) can push the
lattice closer to its breaking strain while not
increasing the $l = m = 2$ quadrupole.

\section{Conclusions and outlook}
\label{concl2}

We have presented a method for calculating the maximum elastic quadrupole
deformation of a relativistic star with a known shear modulus and breaking strain.
We then applied this method to stars whose elastic deformations are
supported by a shear modulus either from the Coulomb lattice of nuclei in the crust, a hadron--quark mixed phase in the core, or crystalline superconducting strange quark matter throughout the star. (In the last case, we have made the requisite changes to the method so that it is valid when the star has a nonzero surface
density and the shear modulus does not vanish at the star's surface.) In all but the strange quark case, we find that the
relativistic quadrupole is suppressed, compared with the standard,
Newtonian Cowling approximation calculation of the quadrupole, at least for
stars with masses of $\gtrsim 1\Msolar$ (corresponding to the observed masses
of neutron stars) and the EOSs we have investigated. These suppressions can be
up to $\sim 4$ in the hybrid case, and
$\sim 6$ in the crustal case. In the strange quark star case, the Newtonian Cowling approximation
calculation slightly underestimates the quadrupole (by tens of percent) for low-to-standard mass stars, but is still an overestimate of $\sim 2$ at higher
masses.

These suppressions strengthen the Horowitz~\cite{Horowitz} argument that
searches for gravitational waves from elastically deformed neutron stars
supported by crustal stresses are biased towards lower-mass stars. The same argument also
applies to strange quark stars, though there the suppressions with increasing mass are less
severe (and the maximum quadrupoles are all considerably larger). However,
this argument does not apply to quadrupole deformations of hybrid stars, since
the increase in the size of the region of mixed phase with increasing mass
dominates the various suppressions.

Our results also imply that many of the
previous calculations of elastic quadrupoles (e.g.,~\cite{Lin, Haskelletal,
KS, UCB, HJA}) will need their results revised downwards. (While we find much larger maximum quadrupoles for solid strange quark stars than did Lin~\cite{Lin}, this is only because we assume a breaking
strain $10$ times that assumed by Lin. If we take the same $10^{-2}$ breaking strain as does Lin, then we find a suppression of a factor of a few, though this is very likely within the uncertainties of Lin's calculation,
which assumed a uniform density, incompressible star with a uniform shear modulus.)

It is instructive to compare our results with the numbers quoted in Pitkin's review~\cite{Pitkin}. All of these were obtained by Pitkin using scalings given in the aforementioned papers,
sometimes updating to the Horowitz and Kadau~\cite{HK} breaking strain, and provide a good overview of the standard Newtonian predictions.
None of our detailed calculations for maximum crustal quadrupoles approach the
high values Pitkin obtained using UCB's fitting formula (as corrected by
\citet{OwenPRL}). However, our very largest hybrid star quadrupoles are an order of magnitude
above Pitkin's quoted maximum, even if one only assumes a breaking strain of $10^{-2}$, as does Pitkin. Additionally, our estimates for maximum solid quark star quadrupoles ($\sim 10^{44}\text{ g cm}^2$ for $1.4\Msolar$ stars) are considerably larger than the ones quoted by Pitkin (based on a different shear modulus model), even if we reduce them by an order of magnitude due to scaling the breaking strain to Pitkin's $10^{-2}$. In fact, they
are in the same range as those Pitkin quotes for a model for crystalline superconducting hybrid stars (with an optimistic gap parameter $5$ times the one we used for solid quark stars, leading to a shear modulus $\sim 40$ times our shear modulus's maximum value).

Even with the relativistic suppressions, we obtain maximum quadrupole deformations of
$\text{a few}\times10^{42}\text{ g cm}^2$ 
in the hybrid case for a very stiff hadronic EOS, and
$\text{a few}\times10^{41}\text{ g cm}^2$ for more realistic cases. In both
situations, the largest maximum quadrupoles are given by the most massive stars. These
values are proportional to the breaking strain and assume that the Horowitz and Kadau~\cite{HK}
breaking strain of about $0.1$ is applicable to the mixed phase. Such large
quadrupole deformations were previously thought only to be possible for solid
quark stars (see~\cite{OwenPRL, Lin, Haskelletal, KS}), or from crustal deformations in the very
low-mass neutron stars considered by Horowitz~\cite{Horowitz}.  These large
deformations (corresponding to fiducial ellipticities of
$\text{a few}\times10^{-3}$ in the extreme case, and
$\sim 5\times10^{-4}$ in a more realistic case)
would be able to be
detected by current LIGO searches for gravitational waves from certain known
neutron stars~\cite{LIGO_psrs2010, LIGO_CasA, LIGO_Vela}. (However, we must note that there is no reason to assume that
such isolated stars are anywhere near maximally strained, even neglecting the
uncertainties in the description of their interiors.)

The prospects for
crustal quadrupoles are now somewhat less optimistic, and definitely favor
lower-mass stars. However, for a canonical $1.4\Msolar$ neutron star, we find that the maximum
relativistic crustal quadrupole is in the range $\sim\text{(1--6)}\times 10^{39}\text{ g cm}^2$ [corresponding to
fiducial ellipticities of $\sim\text{(1--8)}\times 10^{-6}$], depending on the model used for the crust and the
high-density EOS. (Note that the fully consistent Douchin and Haensel model with its associated high-density EOS yields the lowest numbers. Additionally, there is the possibility of a further reduction of up to $\sim 2$ due to the angle averaging procedure used to obtain the shear modulus.) On the high side, these numbers are consistent
with those given previously for breaking strains of $0.1$ by Horowitz~\cite{HK, Horowitz},\footnote{But recall
that the results from Horowitz~\cite{Horowitz} were obtained using the SLy EOS and crustal composition results, so they
are the same as our Newtonian Cowling approximation SLy predictions, given in Fig.~\ref{Q22s_vs_M_SLy_DH_HJA}, except $\sim7\%$ lower, since Horowitz is using the Horowitz and Hughto~\cite{HH} result for the shear modulus. In the fully relativistic case, one requires a thicker crust than provided by the pure SLy results to obtain values for the maximum quadrupole comparable to those given by Horowitz.} though they are a factor of $\sim 5$ lower than the maximum Pitkin~\cite{Pitkin} obtained using scalings of previous results and the maximum
value given by HJA (scaled to this breaking strain). For stars around $2\Msolar$, the relativistic suppressions
lead to maximum quadrupoles that are nearly an order of magnitude smaller than those for a $1.4\Msolar$ star in the compact SLy case: ${\sim\text{(1--5)}}\times 10^{38}\text{ g cm}^2$ [corresponding to
fiducial ellipticities of $\sim\text{(1--6)}\times 10^{-7}$]; and even in the much less compact LKR$1$ case, there is a suppression of $\sim 5$. Previous Newtonian studies (see Fig.~3 in~\cite{Horowitz}) had only
found suppressions of around a factor of $4$, due to the thinning of the crust and the increase
in Newtonian gravity with increasing mass. It will be interesting to consider further models for the crustal
composition and EOS in this case, particularly the large
suite of crustal models including the pasta phases recently calculated by Newton, Gearheart, and Li~\cite{NGL}. (See~\cite{GNHL} for order-of-magnitude estimates of the maximum quadrupole for these models,
illustrating the sensitive dependence on the slope of the symmetry energy.)

One can also compare these maximum elastic quadrupoles with those generated by an internal magnetic field. Here the values depend,
of course, upon the equation of state, compactness, and---perhaps most crucially---magnetic field topology, as well as the quantity one
chooses to use to measure the magnitude of the magnetic field. But
sticking to order-of-magnitude numbers, and considering a canonical $1.4\Msolar$ neutron star, Frieben and
Rezzolla~\cite{FR} show that a toroidal internal
field of $\sim10^{15}$~G would generate a quadrupole of $\sim 10^{39}$--$10^{40}\text{ g cm}^2$, comparable to the
maxima we find for crustal quadrupoles. Similarly, quadrupoles of $\sim 10^{41}$--$10^{42}\text{ g cm}^2$, around the maxima we find for hybrid
stars, could come from magnetic fields of $\sim 10^{16}$~G, while the maximum quadrupoles of $\sim 10^{44}\text{ g cm}^2$ we find for crystalline strange quark stars could also be
generated by magnetic fields of $\sim 10^{17}$~G, close to the maximum allowed field strength. (But note that these magnetic deformations are all computed for ordinary, purely hadronic neutron stars. Additionally,
the quoted maximum elastic quadrupoles in the hybrid case are attained only for more massive stars than the $1.4\Msolar$
stars for which we are quoting the magnetic deformation results.)
The quoted values for magnetic quadrupoles come from the fits given in Sec.~7 of Frieben and Rezzolla~\cite{FR}, except for the final ones, which are obtained from inspection of their Fig.~5 and Table~3. All these values agree in order of magnitude with the
predictions for the twisted torus topology given by Ciolfi, Ferrari, and Gualtieri~\cite{CFG}, and with many other studies for various topologies cited in Frieben and Rezzolla~\cite{FR}. But note that very recent calculations by Ciolfi and Rezzolla~\cite{CR} show that the magnetic field required to obtain a given quadrupole deformation with the twisted torus topology could be reduced by about an order of magnitude if the toroidal contribution dominates.

One would also like to make relativistic calculations of the maximum energy that could be
stored in an elastic deformation.
This would be useful in properly computing the available
energy for magnetar flares, for instance.
(Using Newtonian scalings,  \citet{CO} estimated that the hybrid case was especially
interesting compared to existing LIGO upper limits for gravitational wave emission from such flares.)
The basic expressions (at least in
the perfect fluid case) appear to be readily available in the literature
(see, e.g.,~\cite{Schutz2,DI}; \cite{ST, Finn} give related results including
elasticity). However, one cannot apply these directly to the crustal and hybrid cases, even in the Newtonian
limit, due to the
distributional nature of the density and pressure perturbations. Specifically, the
sudden change in shear modulus at the phase transitions gives delta functions in the
derivatives of the density and pressure perturbations. Since the energy expressions involve
squares of these derivatives, one would have to invoke some sort of regularization procedure, or apply a different
method. Developing appropriate expressions for this case
will be the subject of future work.

Returning to the quadrupoles, one might also want to consider the shape of the deformed star, particularly
in the relativistic case---the ellipticity is already only a rough indicator of the shape of the
deformation in the Newtonian case---as has now been done in~\cite{NKJ-M_shape}.
But the effects of the star's magnetic field are surely the most interesting to consider, from
its influence on the lattices that support elastic deformations, to the changes to the boundary conditions at the star's surface from an external magnetic field (particularly for magnetars), to the internal magnetic field's own contribution to the star's deformation.
One might also want to consider the lattice's full elastic modulus tensor in this case, instead of simply assuming a polycrystalline structure and angle averaging to obtain an effective isotropic shear modulus, as was
done here. (And even if one assumes a polycrystalline structure, one could use more involved, sharper
bounds on the shear modulus than the ones considered here---see~\cite{WDO} for a classic review of such bounds.)

\acknowledgments

We wish to thank S.~Bernuzzi, D.~I.~Jones, A.~Maas, R.~O'Shaughnessy, and the anonymous referee for helpful suggestions.
This work was supported 
by NSF grants PHY-0855589 and PHY-1206027, the Eberly research funds of Penn State,  and the DFG SFB/Transregio 7.

\appendix*

\section{Hadron--quark hybrid stars and the binding energy argument}\label{BEarg}

As we mentioned in Sec.~II~C of~\cite{J-MO1}, if the surface tension is large enough, the mixed phase is not locally favored energetically (i.e., at a fixed baryon density), compared to the individual pure phases. (This was first noted by Heiselberg, Pethick, and Staubo~\cite{HPS} and later
discussed by Alford~\emph{et al.}~\cite{Alfordetal_PRD}.) However, as was also noted in~\cite{J-MO1}, the entire region of mixed phase can still be favored due to global energy arguments, especially when one considers the binding energy of the star (for a fixed total baryon number): One expects the stars with the largest binding
energy (i.e., smallest gravitational mass) for a given total baryon number to be favored. In this
calculation, we always compare with a purely
hadronic star. One would expect the Maxwell construction case with a sharp interface between the two
phases to produce more strongly bound stars than the purely hadronic case, given the local energy results presented in~\cite{HPS, Alfordetal_PRD}. However, at least for the EOSs we consider, the Maxwell construction
stars with total baryon numbers up to the total baryon number of the corresponding maximum mass hybrid star only contain hadronic matter.

Specifically, if we compute the gravitational mass of a hybrid star with a given total baryon number, we find that this mass is
smaller (corresponding to a larger binding energy) than that of 
a purely hadronic star constructed with the same hadronic EOS parameters as the hybrid EOS. However,
these mass differences are not very large, only $\sim0.006\Msolar$ in the most extreme case (the most massive stars with the LKR$1$ EOS), and usually
considerably smaller. One thus might be concerned that this conclusion could be reversed if one includes the contributions of the lattice to the EOS, viz., the lattice's energy density and pressure, and the contributions of the surface tension to the
energy density (through the cell energy). Nevertheless, we find that this is not the case.

Indeed, we find that the mixed phase is favored by the binding energy argument for all the EOS parameters we consider,
even for a surface tension as large as $\sigma = 80\text{ MeV fm}^{-2}$, more than twice as large as the surface tensions
favored by recent calculations~\cite{PKR}, and large enough that the mixed phase is not locally energetically favored. In fact, for these surface tensions, the mixed phase stars with the
additions to the EOS from the blobs and lattice energy are even more strongly favored by the binding energy argument than those with no additions. Of course, as we mentioned in~\cite{J-MO1}, the computations of the lattice additions to the EOS
have some uncertainty, in particular due to our approximate treatment of charge screening. However, we do not expect this to change the qualitative results from the binding energy argument, since that the changes in the binding energy from including the lattice and blob contributions to the energy are relatively small
($\lesssim10\%$). Moreover, we expect that more accurate computations of the cell and lattice energy would reduce their contributions.
Indeed, Christiansen and Glendenning~\cite{ChGl97, ChGl00} argue that the mixed phase should \emph{always} be favored, and any calculation that predicts otherwise must be incomplete or
using inapplicable input parameters.

We now describe the specifics of the binding energy calculation. We calculate the mass differences by first computing the total baryon number as a function of mass for the
purely hadronic stars and then using bisection to locate the hybrid star with the same total baryon number.
However, as noted by Haensel and Pr{\'o}szy{\'n}ski~\cite{HaPr}, the standard method of logarithmic
interpolation of an EOS table is insufficiently accurate to allow one to compute the gravitational masses and baryon numbers with the accuracy we need. One must, instead, use a thermodynamically consistent method of interpolation---i.e., one for which the first law of thermodynamics is satisfied exactly. And, indeed, if we use the standard logarithmic interpolation,
we find that the additions to the EOS have a much larger effect on the binding energy differences, and
the mixed phase is only favored by the binding energy argument for higher masses, if at all.

\begin{table*}
\begin{tabular}{cccccccccc}
\hline\hline
 & $\sigma$ & interpolation & EOS additions & $M_\mathrm{max}$ & $M^\mathrm{hybrid}_\mathrm{min}$ & $R^\mathrm{hybrid}_\mathrm{max}/R$ & $\cC_\mathrm{max}$ & densest\\
 & $(\text{MeV fm}^{-2})$ & & & $(\Msolar)$ & $(\Msolar)$ & $(\%)$ & & hybrid phase\\
\hline
\multirow{3}{*}{Hy1} & -- & log & none & $2.057$ & $1.747$ & $57.7$ & $0.484$ & Q, $d = 1.03$\\
& -- & HaPr & none & $2.047$ & $1.743$ & $57.6$ & $0.484$ & Q, $d = 1.05$\\
& $80$ & HaPr & blob + lattice & $2.040$ & $1.742$ & $57.4$ & $0.483$ & Q, $d = 1.03$\\
\hline
\multirow{3}{*}{Hy$1'$} & -- & log & none & $1.974$ & $1.377$ & $69.0$ & $0.476$ & H, $d = 1.30$\\
& -- & HaPr & none & $1.963$ & $1.375$ & $69.0$ & $0.476$ & H, $d = 1.24$\\
& $80$ & HaPr & blob + lattice & $1.955$ & $1.375$ & $68.9$ & $0.475$ & H, $d = 1.27$\\
\hline
\multirow{3}{*}{LKR1} & -- & log & none & $1.955$ & $1.096$ & $72.5$ & $0.433$ & H, $d = 3.00$\\
& -- & HaPr & none & $1.948$ & $1.098$ & $72.4$ & $0.433$ & H, $d = 3.00$\\
& $80$ & HaPr & blob + lattice & $1.935$ & $1.098$ & $72.3$ & $0.431$ & H, $d = 3.00$\\
\hline
\multirow{3}{*}{generic} & -- & log & none & $1.986$ & $1.878$ & $44.0$ & $0.500$ & Q, $d = 2.10$\\
& -- & HaPr & none & $1.974$ & $1.869$ & $43.8$ & $0.500$ & Q, $d = 2.12$\\
& $80$ & HaPr & blob + lattice & $1.971$ & $1.869$ & $43.4$ & $0.499$ & Q, $d = 2.15$\\
\hline
\multirow{3}{*}{generic$'$} & -- & log & none & $1.974$ & $1.534$ & $65.9$ & $0.515$ & Q, $d = 1.36$\\
& -- & HaPr & none & $1.963$ & $1.528$ & $65.9$ & $0.515$ & Q, $d = 1.36$\\
& $80$ & HaPr & blob + lattice & $1.959$ & $1.528$ & $65.8$ & $0.514$ & Q, $d = 1.38$\\
\hline\hline
\end{tabular}
\caption{\label{mixed_phase_table} Properties of stable stars constructed with the EOSs from~\cite{J-MO1}
(except for the different Hy$1$ ``flavors''), showing the effects of the interpolation and the additions to the EOS. 
In the ``interpolation'' column, ``log'' denotes the standard logarithmic
interpolation of the EOS table (used to obtain the values for stellar quantities given in Table~I of~\cite{J-MO1}, which we repeat here, with the small corrections from the erratum), while ``HaPr'' denotes the Haensel-Pr{\'o}szy{\'n}ski thermodynamically consistent interpolation. In the ``EOS additions'' column, ``blob + lattice'' denotes the case where we have included the blob and lattice energy
densities and lattice pressure in the EOS. The other columns are the same as in Table~I of~\cite{J-MO1}.  Explicitly, $M^\mathrm{hybrid}_\mathrm{min}$ gives the masses of stars that first contain hybrid matter (using the binding energy argument); $R^\mathrm{hybrid}_\mathrm{max}/R$ denotes the maximum radius fraction occupied by hybrid matter
(i.e., the radius fraction for the maximum mass star); and $\cC_\mathrm{max}$
denotes the maximum compactness ($2GM/Rc^2$) of a star. We also give the
composition of the rare phase (``Q'' stands for quark and ``H'' for hadronic) and the
dimension of the lattice at the center of the maximum mass star. (Note that is often necessary to locate the maximum mass with more than its given accuracy to obtain $R^\mathrm{hybrid}_\mathrm{max}/R$, $\cC_\mathrm{max}$, and the dimension of the densest hybrid phase to their given accuracy, as discussed in the erratum to~\cite{J-MO1}.)
}
\end{table*}

Haensel and Pr{\'o}szy{\'n}ski~\cite{HaPr} provide such a thermodynamically consistent method of
interpolation in their Sec.~IIc, which we use to perform the binding energy calculation. There is an alternative expression for the baryon number
density as a function of radius given in Eq.~(6) of Haensel and Potekhin~\cite{HP} (also obtained using
the first law of thermodynamics), but we find the Haensel-Pr{\'o}szy{\'n}ski interpolation to be
preferable, in our experiments. Specifically, we have checked that our qualitative conclusions
remain unchanged if use the EOS output on a finer mesh of baryon number densities (with half the spacing
of the original mesh) and have found that
the results of the Haensel-Pr{\'o}szy{\'n}ski interpolation are less sensitive to changes in the mesh on 
which the EOS table is output than the
Haensel-Potekhin version. We interpret this as indicating that the
Haensel-Pr{\'o}szy{\'n}ski version is more reliable, at least for our situation. (There is also a more involved thermodynamically consistent interpolation
method due to Swesty~\cite{Swesty}, but we have not experimented with this.)

\begin{figure}[htb]
\begin{center}
\epsfig{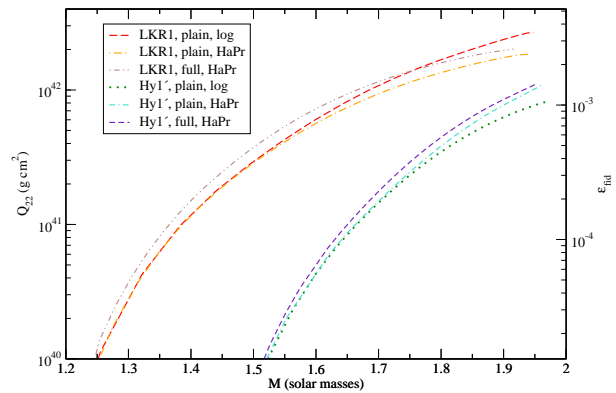}
\end{center}
\caption{\label{Q22_vs_M_EOS_additions} The full relativistic maximum quadrupole deformations (and fiducial ellipticity) of hybrid
stars versus mass, for the LKR$1$ and Hy$1'$ EOSs from~\cite{J-MO1}, with a surface tension of $\sigma = 80\text{ MeV fm}^{-2}$, showing the
effects of the different interpolation methods and EOS additions. Here, ``plain'' denotes no additions, while ``full'' denotes both blob and lattice additions. Similarly, ``log'' and ``HaPr'' denote the standard logarithmic
and Haensel-Pr{\'o}szy{\'n}ski interpolation, respectively.
Note that many of the curves end somewhat short of the maximum mass, due to difficulties with the numerics.}
\end{figure}

We show the differences
in the final stellar quantities calculated using the logarithmic and Haensel-Pr{\'o}szy{\'n}ski interpolation for 
the case of no EOS additions, as well as the effects of the EOS additions with the Haensel-Pr{\'o}szy{\'n}ski interpolation in Table~\ref{mixed_phase_table}. (The $\lesssim0.5\%$ differences in the maximum mass
due to the different methods of interpolation are in line with the differences found by Haensel and Pr{\'o}szy{\'n}ski~\cite{HaPr}, though they find an increase in the maximum mass,
while we only find decreases.) Additionally, including the EOS additions and changing the interpolation also has an effect on the maximum quadrupoles (at most $\sim40\%$), illustrated in Fig.~\ref{Q22_vs_M_EOS_additions} for the two EOSs that yield the largest quadrupoles.

The EOS additions and Haensel-Pr{\'o}szy{\'n}ski interpolation both reduce the maximum mass, compared to the plain logarithmic
interpolation results. Thus, EOSs that already have a low maximum mass (particularly LKR$1$) with the
logarithmic interpolation and no additions may no longer be
consistent within $1\sigma$ with observations of massive neutron stars when using the Haensel-Pr{\'o}szy{\'n}ski interpolation and including the additions. Indeed, these EOSs were designed
to be compatible with the Demorest~\emph{et al.}\ observation of a $1.97\pm0.04\Msolar$ neutron star~\cite{Demorestetal}, so some of them (again, particularly LKR$1$) are not compatible within $1\sigma$ with the very recent observation of a $2.01\pm0.04\Msolar$ neutron star by Antoniadis~\emph{et al.}~\cite{Antoniadisetal}, even with no additions and the logarithmic interpolation. Nevertheless, all of them are still compatible within $2\sigma$, even with the additions and Haensel-Pr{\'o}szy{\'n}ski interpolation. It is also worth pointing out that the Antoniadis~\emph{et al.}\ measurement is less clean than the Demorest~\emph{et al.}\ measurement, as it relies on some modeling of white dwarf atmospheres, not just geometrical considerations.

Additionally, one can easily obtain $1\sigma$ compatibility with the Antoniadis~\emph{et al.}\ measurement with a slight modification of the EOS parameters. For instance, for the LKR$1$ EOS, changing
the QCD coupling constant $\alpha_s$ from $0.6$ to $0.625$ increases the maximum mass to
$2.004\Msolar$ with the logarithmic interpolation and no additions and to $1.984\Msolar$ with the
Haensel-Pr{\'o}szy{\'n}ski interpolation and additions (with a surface tension of $\sigma = 80\text{ MeV fm}^{-2}$), while only decreasing the maximum quadrupoles by
$\lesssim30\%$ for the largest masses

Finally, we describe exactly how we obtain the lattice contributions to the EOS. We compute the lattice and cell energy density [$(E_\text{cell} + W)/\Omega$] using Eqs.~(2) and~(14) in~\cite{J-MO1} and the electrostatic pressure contribution by
multiplying that paper's Eq.~(20) by $d/3$ to account for the angle-averaged
anisotropy ($d$ is the dimension of the lattice). We have also experimented with adding in the isotropic contribution to the pressure from changing the cell energy and blob's charge, given by $-(E_\text{cell}+2W)/\Omega$. We found that this addition does not change the qualitative conclusions,
and, indeed, makes the mixed phase even more strongly favored, giving some indication of the robustness
of the calculation.

\bibliography{paper2}

\end{document}